\pdfoutput=1
\documentclass[12pt,a4paper]{article}

\usepackage{ifthen}
\newboolean{pdflatex}
\setboolean{pdflatex}{true}

\newboolean{articletitles}
\setboolean{articletitles}{true}

\newboolean{uprightparticles}
\setboolean{uprightparticles}{false}

\def\paperauthors{LHCb collaboration}
\def\paperasciititle{Discovery of an unexpectedly light and narrow beauty-strange state}
\def\papertitle{Discovery of an unexpectedly light and narrow beauty-strange state}
\def\paperkeywords{{High Energy Physics}, {LHCb}}
\def\papercopyright{\the\year\ CERN for the benefit of the LHCb collaboration}
\def\paperlicence{CC BY 4.0 licence}
\def\paperlicenceurl{https://creativecommons.org/licenses/by/4.0/}

\newif\ifEnableSectionTOCLinks
\EnableSectionTOCLinksfalse % deactivated

\usepackage[top=1in, bottom=1.25in, left=1in, right=1in]{geometry}

\usepackage{microtype}
\usepackage{lineno}  % for line numbering during review
\usepackage{xspace} % To avoid problems with missing or double spaces after
\usepackage{caption} %these three command get the figure and table captions automatically small

\usepackage{graphicx}  % to include figures (can also use other packages)
\usepackage{color}
\usepackage{colortbl}
\graphicspath{{./figs/}} % Make Latex search fig subdir for figures
\usepackage{amsmath} % Adds a large collection of math symbols
\usepackage{amssymb}
\usepackage{amsfonts}
\usepackage{upgreek} % Adds in support for greek letters in roman typeset

\newcommand*\patchAmsMathEnvironmentForLineno[1]{%
\expandafter\let\csname old#1\expandafter\endcsname\csname #1\endcsname
\expandafter\let\csname oldend#1\expandafter\endcsname\csname
end#1\endcsname
 \renewenvironment{#1}%
   {\linenomath\csname old#1\endcsname}%
   {\csname oldend#1\endcsname\endlinenomath}%
}
\newcommand*\patchBothAmsMathEnvironmentsForLineno[1]{%
  \patchAmsMathEnvironmentForLineno{#1}%
  \patchAmsMathEnvironmentForLineno{#1*}%
}
\AtBeginDocument{%
\patchBothAmsMathEnvironmentsForLineno{equation}%
\patchBothAmsMathEnvironmentsForLineno{align}%
\patchBothAmsMathEnvironmentsForLineno{flalign}%
\patchBothAmsMathEnvironmentsForLineno{alignat}%
\patchBothAmsMathEnvironmentsForLineno{gather}%
\patchBothAmsMathEnvironmentsForLineno{multline}%
\patchBothAmsMathEnvironmentsForLineno{eqnarray}%
}

\usepackage[pdftex,
            pdfauthor={\paperauthors},
            pdftitle={\paperasciititle},
            pdfkeywords={\paperkeywords}]{hyperref}
\usepackage{hyperxmp}
\hypersetup{
    pdfcopyright={Copyright (C) \papercopyright},
    pdflicenseurl={\paperlicenceurl}
}
\usepackage[colorinlistoftodos,textsize=scriptsize]{todonotes}

\usepackage[bottom,flushmargin,hang,multiple]{footmisc}

\usepackage[all]{hypcap} % Internal hyperlinks to floats.

\usepackage{xspace} 
\usepackage{upgreek}

\def\lhcb   {\mbox{LHCb}\xspace}

\def\MagUp {\mbox{\em Mag\kern -0.05em Up}\xspace}

\ifthenelse{\boolean{uprightparticles}}%
{
 
 \def\Pgamma      {\ensuremath{\upgamma}\xspace}

 \def\Pmu         {\ensuremath{\upmu}\xspace}

 \def\Ppi         {\ensuremath{\uppi}\xspace}

 \def\Pphi        {\ensuremath{\upphi}\xspace}

 \def\Ppsi        {\ensuremath{\uppsi}\xspace}

 \def\PDelta      {\ensuremath{\Delta}\xspace}                 
 \def\PXi         {\ensuremath{\Xi}\xspace}                 
 \def\PLambda     {\ensuremath{\Lambda}\xspace}                 
 \def\PSigma      {\ensuremath{\Sigma}\xspace}                 
 \def\POmega      {\ensuremath{\Omega}\xspace}                 
 \def\PUpsilon    {\ensuremath{\Upsilon}\xspace}
 \let\oldPi\Pi
 \def\PPi         {\ensuremath{\oldPi}\xspace}

 \def\PB      {\ensuremath{\mathrm{B}}\xspace}                 
 \def\PD      {\ensuremath{\mathrm{D}}\xspace}                 
 \def\PJ      {\ensuremath{\mathrm{J}}\xspace}                 
 \def\PK      {\ensuremath{\mathrm{K}}\xspace}                 
 \def\Pb      {\ensuremath{\mathrm{b}}\xspace}                 
 \def\Pc      {\ensuremath{\mathrm{c}}\xspace}

 \def\Pp      {\ensuremath{\mathrm{p}}\xspace}                 
 \def\Pq      {\ensuremath{\mathrm{q}}\xspace}                 
 \def\Ps      {\ensuremath{\mathrm{s}}\xspace}

 \def\thebaroffset{0.0em}
}
{
 
 \def\Pgamma      {\ensuremath{\gamma}\xspace}

 \def\Pmu         {\ensuremath{\mu}\xspace}

 \def\Ppi         {\ensuremath{\pi}\xspace}

 \def\Pphi        {\ensuremath{\phi}\xspace}

 \def\Ppsi        {\ensuremath{\psi}\xspace}                 
                  
 \mathchardef\PDelta="7101
 \mathchardef\PXi="7104
 \mathchardef\PLambda="7103
 \mathchardef\PSigma="7106
 \mathchardef\POmega="710A
 \mathchardef\PUpsilon="7107
 \mathchardef\PPi="7105
 \def\PB      {\ensuremath{B}\xspace}                 
 \def\PD      {\ensuremath{D}\xspace}                 
 \def\PJ      {\ensuremath{J}\xspace}                 
 \def\PK      {\ensuremath{K}\xspace}                 
 \def\Pb      {\ensuremath{b}\xspace}                 
 \def\Pc      {\ensuremath{c}\xspace}

 \def\Pp      {\ensuremath{p}\xspace}                 
 \def\Pq      {\ensuremath{q}\xspace}                 
 \def\Ps      {\ensuremath{s}\xspace}

 \def\thebaroffset{0.18em}
}
\newcommand{\offsetoverline}[2][\thebaroffset]{\kern #1\overline{\kern -#1 #2}}%

\makeatletter
\ifcase \@ptsize \relax% 10pt
  \newcommand{\miniscule}{\@setfontsize\miniscule{4}{5}}% \tiny: 5/6
\or% 11pt
  \newcommand{\miniscule}{\@setfontsize\miniscule{5}{6}}% \tiny: 6/7
\or% 12pt
  \newcommand{\miniscule}{\@setfontsize\miniscule{5}{6}}% \tiny: 6/7
\fi
\makeatother

\DeclareRobustCommand{\optbar}[1]{\shortstack{{\miniscule (\rule[.5ex]{1.25em}{.18mm})}
  \\ [-.7ex] $#1$}}

\def\mumu       {{\ensuremath{\Pmu^+\Pmu^-}}\xspace}

\def\g      {{\ensuremath{\Pgamma}}\xspace}

\def\quark     {{\ensuremath{\Pq}}\xspace}
\def\quarkbar  {{\ensuremath{\overline \quark}}\xspace}

\def\squark    {{\ensuremath{\Ps}}\xspace}

\def\cquark    {{\ensuremath{\Pc}}\xspace}

\def\bquark    {{\ensuremath{\Pb}}\xspace}

\def\pion   {{\ensuremath{\Ppi}}\xspace}
\def\piz    {{\ensuremath{\pion^0}}\xspace}
\def\pip    {{\ensuremath{\pion^+}}\xspace}
\def\pim    {{\ensuremath{\pion^-}}\xspace}
\def\pipm   {{\ensuremath{\pion^\pm}}\xspace}

\def\kaon    {{\ensuremath{\PK}}\xspace}
\def\KorKbar {\kern \thebaroffset\optbar{\kern -\thebaroffset \PK}{}\xspace}

\def\Kp      {{\ensuremath{\kaon^+}}\xspace}
\def\Km      {{\ensuremath{\kaon^-}}\xspace}

\def\KS      {{\ensuremath{\kaon^0_{\mathrm{S}}}}\xspace}

\def\D       {{\ensuremath{\PD}}\xspace}

\def\DorDbar {\kern \thebaroffset\optbar{\kern -\thebaroffset \PD}\xspace}
\def\Dz      {{\ensuremath{\D^0}}\xspace}

\def\Dp      {{\ensuremath{\D^+}}\xspace}
\def\Dm      {{\ensuremath{\D^-}}\xspace}

\def\DpDm    {\ensuremath{\Dp {\kern -0.16em \Dm}}\xspace}

\def\Dstarp  {{\ensuremath{\D^{*+}}}\xspace}

\def\Ds      {{\ensuremath{\D^+_\squark}}\xspace}

\def\Dsm     {{\ensuremath{\D^-_\squark}}\xspace}

\def\Dss     {{\ensuremath{\D^{*+}_\squark}}\xspace}

\def\B       {{\ensuremath{\PB}}\xspace}

\def\BorBbar {\kern \thebaroffset\optbar{\kern -\thebaroffset \PB}\xspace}

\def\Bd      {{\ensuremath{\B^0}}\xspace}

\def\BdorBdbar {\kern \thebaroffset\optbar{\kern -\thebaroffset \Bd}\xspace}

\def\Bs      {{\ensuremath{\B^0_\squark}}\xspace}

\def\BsorBsbar {\kern \thebaroffset\optbar{\kern -\thebaroffset \Bs}\xspace}

\def\jpsi     {{\ensuremath{{\PJ\mskip -3mu/\mskip -2mu\Ppsi}}}\xspace}

\def\Y#1S{\ensuremath{\PUpsilon{(#1S)}}\xspace}

\def\proton      {{\ensuremath{\Pp}}\xspace}

\def\Lz          {{\ensuremath{\PLambda}}\xspace}

\def\LorLbar     {\kern \thebaroffset\optbar{\kern -\thebaroffset \PLambda}\xspace}

\def\Sigmares    {{\ensuremath{\PSigma}}\xspace}

\def\Xires       {{\ensuremath{\PXi}}\xspace}

\def\Omegares    {{\ensuremath{\POmega}}\xspace}

\def\Lc          {{\ensuremath{\Lz^+_\cquark}}\xspace}

\def\Sigmac       {{\ensuremath{\Sigmares_\cquark}}\xspace}

\def\Xicp        {{\ensuremath{\Xires^+_\cquark}}\xspace}

\def\Xicc        {{\ensuremath{\Xires_{\cquark\cquark}}}\xspace}

\def\Xiccp       {{\ensuremath{\Xires^+_{\cquark\cquark}}}\xspace}
\def\Xiccpp      {{\ensuremath{\Xires^{++}_{\cquark\cquark}}}\xspace}

\def\Omegacc     {{\ensuremath{\Omegares^+_{\cquark\cquark}}}\xspace}

\newcommand{\decay}[2]{\ensuremath{\mathinner{#1\!\to #2}}\xspace}

\def\to                 {\ensuremath{\rightarrow}\xspace}

\def\AT#1     {\ensuremath{A_{\mathrm{T}}^{#1}}\xspace}           % 2

\def\C#1      {\ensuremath{\mathcal{C}_{#1}}\xspace}                       % 9
\def\Cp#1     {\ensuremath{\mathcal{C}_{#1}^{'}}\xspace}                    % 7
\def\Ceff#1   {\ensuremath{\mathcal{C}_{#1}^{\mathrm{(eff)}}}\xspace}        % 9  
\def\Cpeff#1  {\ensuremath{\mathcal{C}_{#1}^{'\mathrm{(eff)}}}\xspace}       % 7
\def\Ope#1    {\ensuremath{\mathcal{O}_{#1}}\xspace}                       % 2
\def\Opep#1   {\ensuremath{\mathcal{O}_{#1}^{'}}\xspace}                    % 7

\newcommand{\aunit}[1]{\ensuremath{\text{\,#1}}}       
\newcommand{\tev}{\aunit{Te\kern -0.1em V}\xspace}
\newcommand{\gev}{\aunit{Ge\kern -0.1em V}\xspace}
\newcommand{\mev}{\aunit{Me\kern -0.1em V}\xspace}
\newcommand{\kev}{\aunit{ke\kern -0.1em V}\xspace}
\newcommand{\ev}{\aunit{e\kern -0.1em V}\xspace}
 
\newcommand{\mevc}{\ensuremath{\aunit{Me\kern -0.1em V\!/}c}\xspace}
\newcommand{\gevc}{\ensuremath{\aunit{Ge\kern -0.1em V\!/}c}\xspace}
\newcommand{\mevcc}{\ensuremath{\aunit{Me\kern -0.1em V\!/}c^2}\xspace}
\newcommand{\gevcc}{\ensuremath{\aunit{Ge\kern -0.1em V\!/}c^2}\xspace}
\def\fb   {\ensuremath{\aunit{fb}}\xspace}
\def\invfb   {\ensuremath{\fb^{-1}}\xspace}

\newcommand{\stat}{\aunit{(stat)}\xspace}
\newcommand{\syst}{\aunit{(syst)}\xspace}

\def\gsim{{~\raise.15em\hbox{$>$}\kern-.85em
          \lower.35em\hbox{$\sim$}~}\xspace}
\def\lsim{{~\raise.15em\hbox{$<$}\kern-.85em
          \lower.35em\hbox{$\sim$}~}\xspace}

\def\pt         {\ensuremath{p_{\mathrm{T}}}\xspace}

\def\evtgen     {\mbox{\textsc{EvtGen}}\xspace}

\def\geant      {\mbox{\textsc{Geant4}}\xspace}

\def\photos     {\mbox{\textsc{Photos}}\xspace}

\def\pythia     {\mbox{\textsc{Pythia}}\xspace}

\def\tell1  {TELL1\xspace}
\def\ukl1   {UKL1\xspace}

\newcommand{\ie}{\mbox{\itshape i.e.}\xspace}

\newcommand{\lhcborcid}[1]{\href{https://orcid.org/#1}{\hspace*{0.1em}\raisebox{-0.45ex}{\includegraphics[width=1em]{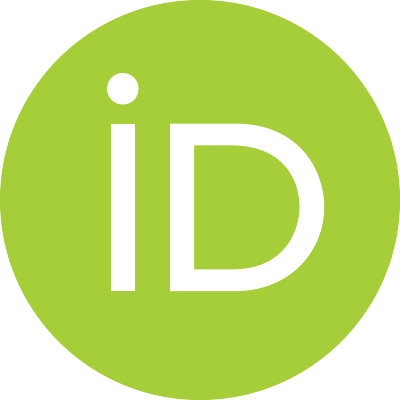}}}}

\hypersetup{
  colorlinks   = true, %Colours links instead of ugly boxes
  urlcolor     = blue, %Colour for external hyperlinks
  linkcolor    = blue, %Colour of internal links
  citecolor    = red   %Colour of citations
}

\ifEnableSectionTOCLinks
    \usepackage[explicit]{titlesec} % to change headings
    
    \let\oldcontentsline\contentsline
    \renewcommand

    \titleformat{\section}{\normalfont\Large\bf}{\hyperlink{tocsection.\thesection}{{\thesection} \parbox[t]{\dimexpr\textwidth-1pc}{#1}}}{1pc}{}

    \titleformat{\subsection}{\normalfont\bf}{\hyperlink{tocsubsection.\thesubsection}{{\thesubsection} \parbox[t]{\dimexpr\textwidth-1pc}{#1}}}{1pc}{}

    \titleformat{name=\section,numberless}[display]{}{}{0pt}{\normalfont\Huge\bfseries #1}
\fi

\usepackage{cite} % Allows for ranges in citations
\usepackage{LHCb/mciteplus}
\def\Dsstz      {{\ensuremath{\D_{\squark 0}^{*}(2317)^+}}\xspace}
\def\Dsstone      {{\ensuremath{\D_{\squark 1}(2460)^+}}\xspace}

\def\Bsstz      {{\ensuremath{\B_{\squark 0}^{*0}}}\xspace}

\def\Bsstznum      {{\ensuremath{\B_{\squark 0}^{*}(5700)^0}}\xspace}
\def\Bsst      {{\ensuremath{\B_{\squark}^{*0}}}\xspace}

\DeclareRobustCommand{\optbarbis}[1]{\shortstack{{\miniscule (\rule[.5ex]{0.6em}{.18mm})}
  \\ [-.5ex] $#1$}}

\def\thebaroffsetbis{-0.35em}
\def\thebaroffsetter{0.15em}
\def\qorqbar {\kern \thebaroffsetbis\optbarbis{\kern -\thebaroffsetter \quark}\xspace}
\def\qorqbarbis {\optbar{\kern -\thebaroffset \quark}\xspace}

\usepackage{hyperref}

\begin{document}

%%%%%%%%%%%%%%%%%%%%%%%%%
%%%%% Title     %%%%%%%%%
%%%%%%%%%%%%%%%%%%%%%%%%%
\renewcommand{\thefootnote}{\fnsymbol{footnote}}
\setcounter{footnote}{1}

% ===============================================================================
% Purpose: LHCb-PAPER journal paper title page template
% Author: 
% Created on: 2010-09-25
% ===============================================================================

%%%%%%%%%%%%%%%%%%%%%%%%%
%%%%%  TITLE PAGE  %%%%%%
%%%%%%%%%%%%%%%%%%%%%%%%%
\begin{titlepage}
\pagenumbering{roman}

% Header ---------------------------------------------------
\vspace*{-1.5cm}
\centerline{\large EUROPEAN ORGANIZATION FOR NUCLEAR RESEARCH (CERN)}
\vspace*{1.5cm}
\noindent
\begin{tabular*}{\linewidth}{lc@{\extracolsep{\fill}}r@{\extracolsep{0pt}}}
\ifthenelse{\boolean{pdflatex}}% Logo format choice
{\vspace*{-1.5cm}\mbox{\!\!\!\includegraphics[width=.14\textwidth]{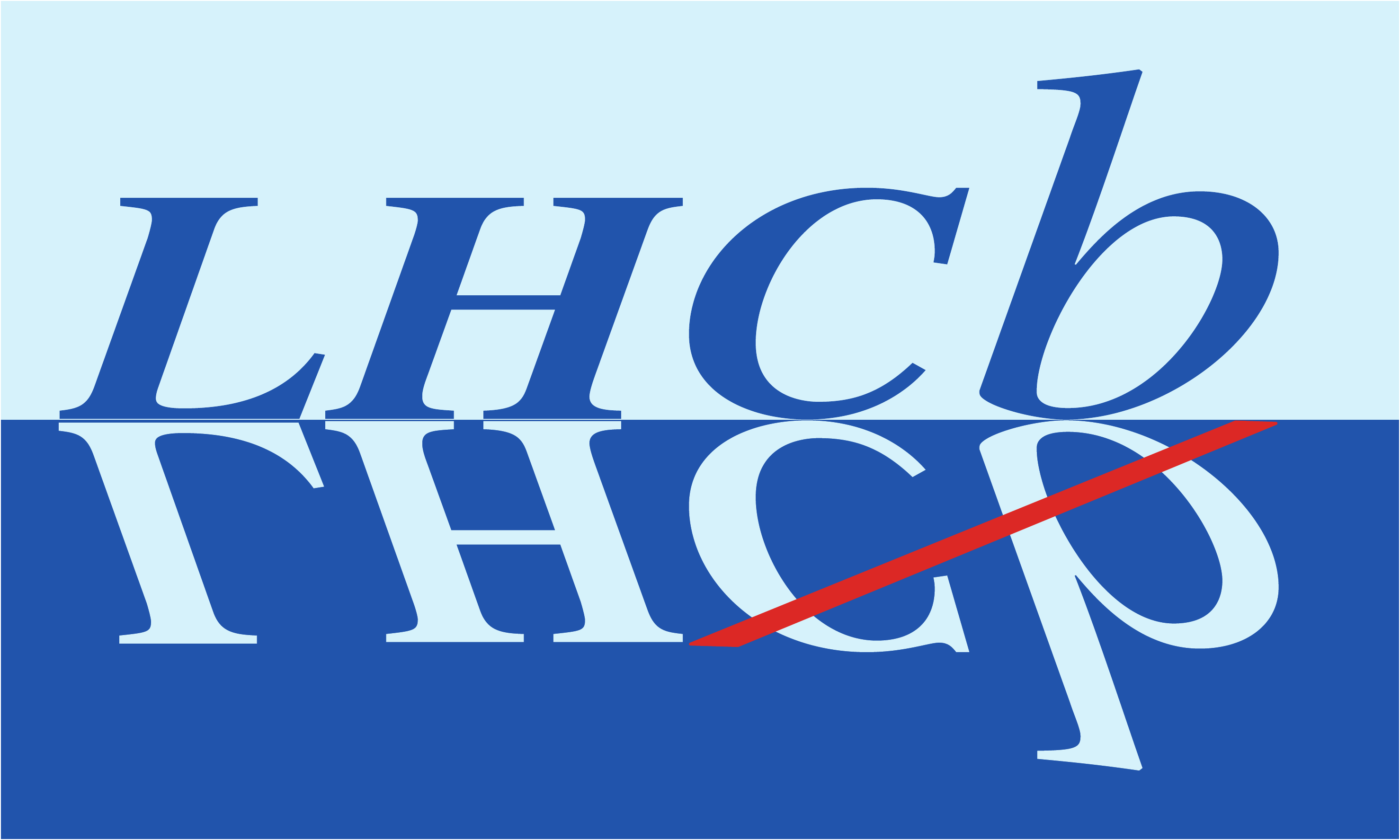}} & &}%
{\vspace*{-1.2cm}\mbox{\!\!\!\includegraphics[width=.12\textwidth]{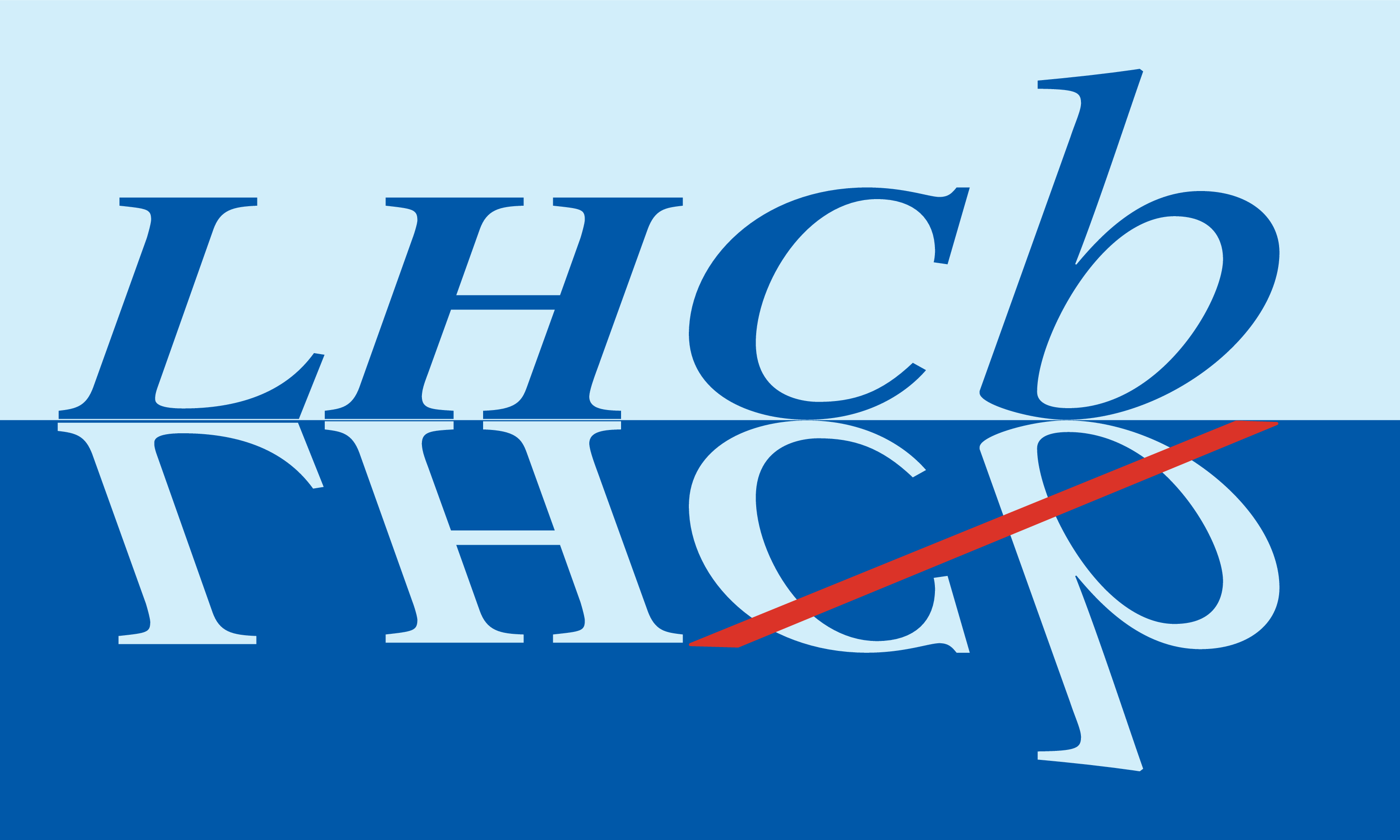}} & &}%
\\
 & & CERN-EP-2026-250 \\  % ID 
 & & LHCb-PAPER-2026-021 \\  % ID 
 & & September 21, 2026 \\ % Date - Can also hardwire e.g.: 23 March 2010
 & & \\
% not in paper \hline
\end{tabular*}

\vspace*{2.5cm}

% Title --------------------------------------------------
{\normalfont\bfseries\boldmath\huge
\begin{center}
% DO NOT EDIT HERE. Instead edit macro in main.tex to keep metadata correct
  \papertitle 
\end{center}
}

\vspace*{0.5cm}

% Authors -------------------------------------------------
\begin{center}
%In the footnote, replace 'paper' by 'Letter' in case of submission to PRL or PLB 
% Edit macro in main.tex to keep metadata correct
\paperauthors\footnote{Authors are listed at the end of this paper.}
\end{center}

%\vspace{\fill}
\vspace*{1cm}

% Abstract -----------------------------------------------
\begin{abstract}
  \noindent
  As the essential building blocks of visible matter, hadrons have traditionally been classified by the quark model as mesons composed of quark-antiquark pairs and baryons built from three valence quarks.
  Yet, this classical picture does not account for the intricate chiral dynamics of the strong force and the emergence of exotic multi-quark hadrons.
  While experimental evidence for such unconventional dynamics has surfaced in the charm sector,
  the open-beauty system remains the long-sought frontier for testing the universality of these mechanisms.
  Here the observation of a new resonance in the beauty-strange sector with a global significance exceeding seven standard deviations is reported using proton-proton collision data recorded by the Large Hadron Collider beauty~(LHCb) experiment at the European Organization for Nuclear Research (CERN).
  It exhibits a narrow natural width and a substantial mass deficit compared to the conventional quark model predictions.
  This result marks the first observation of a nonconventional single-beauty hadron, providing crucial insights into the chiral dynamics of the strong interaction and heavy-quark spin symmetry in the exotic domain.

\end{abstract}

\vspace*{0.5cm}

\begin{center}

\end{center}

\vspace{\fill}

{\footnotesize 
% Edit macro in main.tex to keep metadata correct
\centerline{\copyright~\papercopyright. \href{\paperlicenceurl}{\paperlicence}.}}
\vspace*{2mm}

\end{titlepage}

%%%%%%%%%%%%%%%%%%%%%%%%%%%%%%%%
%%%%%  EOD OF TITLE PAGE  %%%%%%
%%%%%%%%%%%%%%%%%%%%%%%%%%%%%%%%

%  empty page follows the title page ----
\newpage
\setcounter{page}{2}
\mbox{~}

\renewcommand{\thefootnote}{\arabic{footnote}}
\setcounter{footnote}{0}

\cleardoublepage

%%%%%%%%%%%%%%%%%%%%%%%%%
%%%%% Main text %%%%%%%%%
%%%%%%%%%%%%%%%%%%%%%%%%%

\pagestyle{plain} % restore page numbers for the main text
\setcounter{page}{1}
\pagenumbering{arabic}

%\linenumbers

%%%%%%%%%%%%%%%%%%%%%%%%%%%%%%%%
%%%%      Introduction
%%%%%%%%%%%%%%%%%%%%%%%%%%%%%%%%

Virtually all visible matter in the universe is composed of hadrons,
which are bound states of quarks and gluons held together by the strong interaction.
The fundamental theory of the strong interaction is Quantum Chromodynamics (QCD) according to which the interaction strength between quarks and gluons decreases as the energy scale of the interaction increases. At high energies, quarks and gluons therefore behave as quasi-free particles~\cite{Gross:1973id,Politzer:1973fx}, providing the theoretical foundation for the successful application of perturbative QCD to hard scattering processes.
However, as energy decreases and the strength of the strong force grows,  QCD exhibits two profound phenomena that shape the properties of the hadronic
world: colour confinement, which binds quarks and gluons into colour-neutral hadrons~\cite{Wilson:1974sk},
and the spontaneous breaking of chiral symmetry, which explains why nucleons are nearly two orders of magnitude heavier than the combined bare masses of their constituent quarks, thereby generating over 98\% of the baryonic mass of the universe~\cite{Nambu:1961tp, Gell-Mann:1968hlm, Roberts:2021nhw}. 
Understanding how the hadronic matter emerges from the complex dynamics of QCD remains a key fundamental challenge in particle physics because of the nonperturbative nature of the theory in this low-energy regime.
Hadron spectroscopy provides critical insights into QCD by mapping the masses and decay properties of hadronic states~\cite{PDG2026,Brambilla:2010cs,Chen:2016qju},
which inherently incorporate both colour confinement and chiral symmetry breaking.

Historically, hadron spectroscopy has been a primary driver of particle physics throughout the 20th century.
The proliferation of hadronic states, once termed the ``particle zoo", found elegant order through the quark model proposed by Gell-Mann,  Zweig and Petermann~\cite{GellMann:1964nj,Zweig:352337,Petermann:1965qlk}, which classified the known hadrons into mesons consisting of quark-antiquark pairs, \quark\quarkbar, and (anti)baryons made of three (anti)quarks, \quark\quark\quark ($\quarkbar\kern 0.05em \quarkbar \kern 0.05em \quarkbar$).
This breakthrough ultimately culminated in the establishment of QCD.
In the subsequent decades, advances in hadron spectroscopy drove the establishment and refinement of the quark-potential model as the standard paradigm for conventional hadrons~\cite{Godfrey:1985xj,Capstick:1986ter}.

A major turning point came in 2003 with the discovery of the $\Dsstz$ and $\Dsstone$ mesons\cite{BaBar:2003oey, CLEO:2003ggt} in the \Ds\piz and \Dss\piz final states, respectively.
Mesons, such as \Ds, composed of a heavy quark (charm $c$) and a light antiquark (strange $\bar s$),  can be grouped into same-parity $P$ doublets, where the spins $J$ of the two states differ by one unit: $J^P = (0^-, 1^-)$ for $L=0$, $(0^+, 1^+)$ and $(1^+, 2^+)$ for $L=1$, and so on, where $L$ is the relative orbital angular momentum between the constituent quarks. The $(0^-, 1^-)$ doublet is identified with the ground states $\Ds$ and $\Dss$, while the $(1^+, 2^+)$ doublet comprises the $\D_{s1}(2536)^+$ and $\D_{s2}(2573)^+$ resonances.
Although the spin and parity of the $\Dsstz$ and $\Dsstone$ mesons are consistent with $0^+$ and $1^+$ respectively~\cite{Belle:2004dpt, Belle:2003guh, BaBar:2004yux}, their identification as the missing $(0^+,1^+)$ \Ds doublet is challenged by their anomalously low masses, approximately 150\mevcc below the masses predicted by quark-potential models~\cite{Godfrey:1985xj,Godfrey:2015dva,Lahde:1999ih,DiPierro:2001dwf}. 
This mass deficit is accompanied by exceptionally narrow widths,
as favoured strong decay modes are energetically forbidden, leaving the states to decay almost exclusively via the 
${\D^{}_\squark}{\kern -0.15 cm ^{(*)+}\piz}$
decay mode, which is severely suppressed by isospin symmetry.

These observations triggered a flurry of theoretical activities~\cite{
Matsuki:2006zoi,Lakhina:2006fy,WooLee:2006kdh,
Koponen:2007nr,Gregory:2010gm,Lang:2015hza,Wurtz:2015mqa,Hudspith:2023loy,Gayer:2024akw, Liu:2012zya,
Bardeen:2003kt,Nowak:2003ra, Colangelo:2005gb,Badalian:2007yr,Colangelo:2012xi,Wang:2015mxa, Alhakami:2020vil,Gandhi:2022nnk,Wang:2007tu,Vishwakarma:2022sly,
Kolomeitsev:2003ac,Barnes:2003dj, Guo:2006fu,Altenbuchinger:2013vwa,Ortega:2016pgg,Albaladejo:2016ztm,Zhou:2020moj,Guo:2021rjv,Ni:2023lvx,Zhang:2024usz, Cleven:2010aw,Fu:2021wde,
Vijande:2007ke,Torres-Rincon:2014ffa,Du:2017zvv,Yang:2022vdb, Kim:2023htt, 
Cheng:2003kg, Maiani:2004vq,Dmitrasinovic:2012zz}, 
pointing to nonconventional interpretations for the structure of this doublet.
They primarily centre around two paradigms, both fundamentally governed by chiral symmetry:
one interprets the $\Dsstz$ and $\Dsstone$ doublet as the $(0^+$,$1^+)$ chiral partners of the ground-state $(0^-,1^-)$ $\Ds$ mesons within the chiral doubling framework~\cite{Bardeen:2003kt, Nowak:2003ra}, 
while the other describes them as exotic hadronic states beyond the conventional $q\bar{q}$  picture~\cite{Olsen:2017bmm, Guo:2017jvc, Karliner:2017qhf, Liu:2019zoy}, such as $D^{(*)}K$ hadronic molecules~\cite{Barnes:2003dj,Guo:2006fu} or compact $c\bar{s}q\bar{q}$ tetraquarks~\cite{Cheng:2003kg, Maiani:2004vq, Dmitrasinovic:2012zz}, with their dynamics governed by heavy-hadron chiral perturbation theory~\cite{Wise:1992hn,Yan:1992gz}.

The investigation of the beauty sector is essential to ultimately determine whether the $\Dsstz$ and $\Dsstone$ anomalies represent an isolated phenomenon~\cite{Lakhina:2006fy} or a universal pattern rooted in the chiral dynamics of QCD.
In the latter scenario, the beauty-strange $0^+$ scalar state is predicted, analogously to its charm counterpart \Dsstz, 
to fall below the threshold formed by the combined $\B$ and $\kaon$ masses.
This low mass closes the isospin-allowed decays into the $\B\kaon$ final states, leaving open only isospin-violating transitions $\Bs \piz$ (via $\piz$-$\eta$ mixing~\cite{Cho:1994zu}) and radiative decays $\Bsst\g$, 
thus yielding an exceptionally narrow resonance~\cite{Bardeen:2003kt}.
Notably, no exotic hadron containing a single $\bquark$ quark has been firmly observed so far
(a narrow structure $X(5568)^\pm$ in the $\Bs\pipm$ spectrum was reported~\cite{D0:2016mwd}, but not confirmed by other experiments~\cite{LHCB-PAPER-2016-029, CMS:2017hfy, CDF:2017dwr, ATLAS:2018udc}).
Therefore, the observation of the beauty counterpart of the $\Dsstz$ state
could offer a unique platform to test heavy-quark flavour symmetry and quantify its breaking effects in the exotic hadron domain.

%%%%%%%%%%%%%%%%%%%%%%%%%%%%%%%%
%%%%      Detector
%%%%%%%%%%%%%%%%%%%%%%%%%%%%%%%%
In this paper, the observation of a new state in the $\Bs \piz$ invariant-mass spectrum is reported, consistent with the expectation of the $\bquark$-quark counterpart of $\Dsstz$.
The study is based on proton–proton ($pp$) collision data collected by the \lhcb detector~\cite{LHCb-DP-2008-001,LHCb-DP-2014-002} at the Large Hadron Collider (LHC) from the year 2011 to 2018 at centre-of-mass energies of 7, 8, and 13\tev, corresponding to an integrated luminosity of 9\invfb. 
The \lhcb detector is a single-arm forward spectrometer designed to study particles containing $b$ or $c$ quarks.
Simulated samples are produced to model the signal characteristics using the standard \lhcb setup.

%%%%%%%%%%%%%%%%%%%%%%%%%%%%%%%%
%%%%       Analysis
%%%%%%%%%%%%%%%%%%%%%%%%%%%%%%%%
%% ======= Reconstruction & Selection
To build the $\Bs \piz$ candidates,
the $\Bs$ mesons are first reconstructed using three decay modes,
$\decay{\Bs}{\Dsm\pip}$, 
$\decay{\Bs}{ \jpsi\phi}$
and $\decay{\Bs}{ \Dsm\pip\pip\pim}$,
where $\Dsm$, $\jpsi$ and $\phi$ are reconstructed as $\Dsm\to \Kp\Km\pim$, $\jpsi\to\mumu$ and $\phi\to\Kp\Km$.
Charge conjugation is implied throughout the paper.
For each decay mode, combinatorial background, arising from random particle associations, is suppressed with single-variable requirements and a subsequent boosted decision-tree~(BDT) classifier~\cite{AdaBoost},
exploiting the distinct kinematic, topological, track and vertex quality, and particle-identification properties of the $\Bs$ decays.
The $\Bs$ candidates are further combined with $\piz$ mesons reconstructed from the $\decay{\piz}{\gamma\gamma}$ decay,
requiring the two photons to be identified as separate clusters in the electromagnetic calorimeter. 
The resulting $\Bs \piz$ candidates are subject to a similar two-stage selection of single-variable requirements followed by a dedicated BDT classifier.  
The selection criteria of $\Bs\piz$ candidates are  optimised for each $\Bs$ decay mode by maximising the figure of merit described in Ref.~\cite{Punzi:2003bu}.
A detailed description of the reconstruction and selection strategy, as well as the detector and simulation frameworks, is provided in \hyperref[sec:methods]{Methods}.

%% ======= Sig & Bkg components
A narrow peaking structure is clearly visible in the $\Bs \piz$ invariant-mass distributions, appearing at a common position across all three $\Bs$ decay modes. The structure remains stable  across different data-taking periods, \lhcb dipole magnet polarities~\cite{LHCb-TDR-001} and selection criteria.
Extensive checks using event mixing, where candidates from independent events are paired to construct an uncorrelated background sample, and sideband control samples show no comparable peak, ruling out potential spurious peaks induced by kinematic selection criteria or combinatorial artifacts.

\begin{figure}[t]
  \centering
  \includegraphics[width=0.6\textwidth]{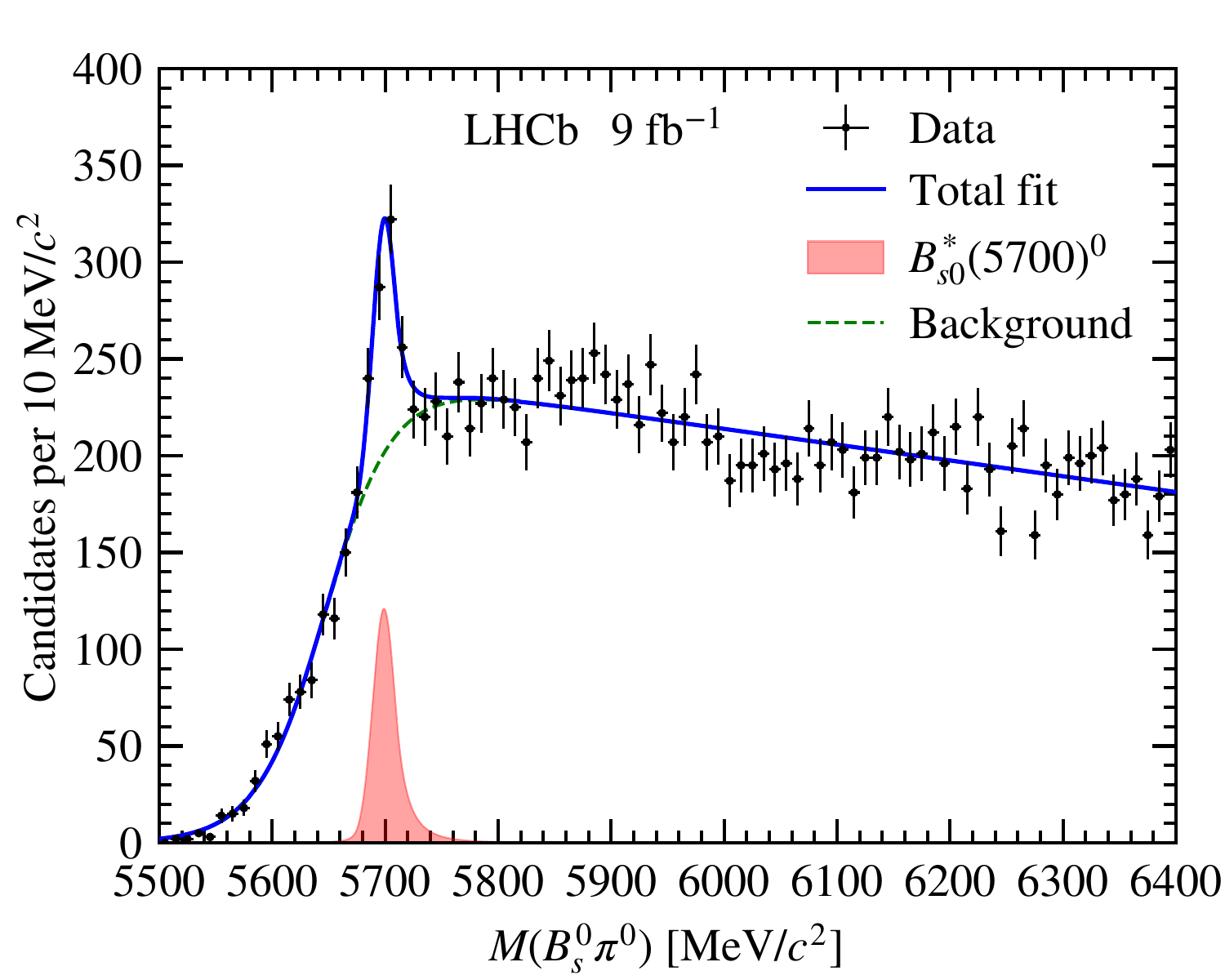}
  \caption{
    Invariant-mass distribution of the $\Bs\piz$ candidates summed over the three $\Bs$ decay modes with the total fit projection overlaid.
  }
  \label{fig:massfit_all}
\end{figure}

A simultaneous extended unbinned maximum-likelihood fit is performed to the $\Bs \piz$ mass distributions of all three decay modes to evaluate the properties of the peak.
The $\Bs \piz$ mass resolution is refined by constraining the reconstructed $\Dsm$, $\jpsi$, $\Bs$ and $\piz$ masses to their world-average values~\cite{PDG2026} and requiring the candidate to originate from the $pp$ interaction point~\cite{Hulsbergen:2005pu}.
The signal component is modelled by a relativistic $S$-wave Breit--Wigner function with the resonance mass and natural width shared across the three decay modes, convolved with a decay-mode dependent detector response function.
The response function, parametrised as a Gaussian kernel with power-law tails, is determined from simulation.
The possible small differences between simulation and data are accounted for by a resolution scale factor determined using the $\decay{\Sigmac(2455)^+}{\Lc\piz}$, $\decay{\Sigmac(2520)^+}{\Lc\piz}$, and $\decay{\Xicp}{\Lc\piz}$ decays, as detailed in \hyperref[sec:methods]{Methods}.
The background contribution is modelled using a sigmoid function modulated by a first-order polynomial following Refs.~\cite{LHCb-PAPER-2011-019,LHCb-PAPER-2013-028,LHCb-PAPER-2018-032}, with its parameters free to vary independently in each decay mode.
Figure~\ref{fig:massfit_all} displays the mass spectrum summed over all decay modes, along with the corresponding total fit projection.
The Kolmogorov–Smirnov test~\cite{Kolmogorov1933} yields $p$-values exceeding $40\%$ for all three decay modes,
indicating excellent agreement between data and the fit model, without overfitting.
The individual mass distributions and corresponding fit results for each decay mode are provided in \hyperref[sec:methods]{Methods}, Fig.~\ref{fig:massfit_3modes}.

\begin{figure}[t]
  \centering
  \includegraphics[width=0.6\textwidth]{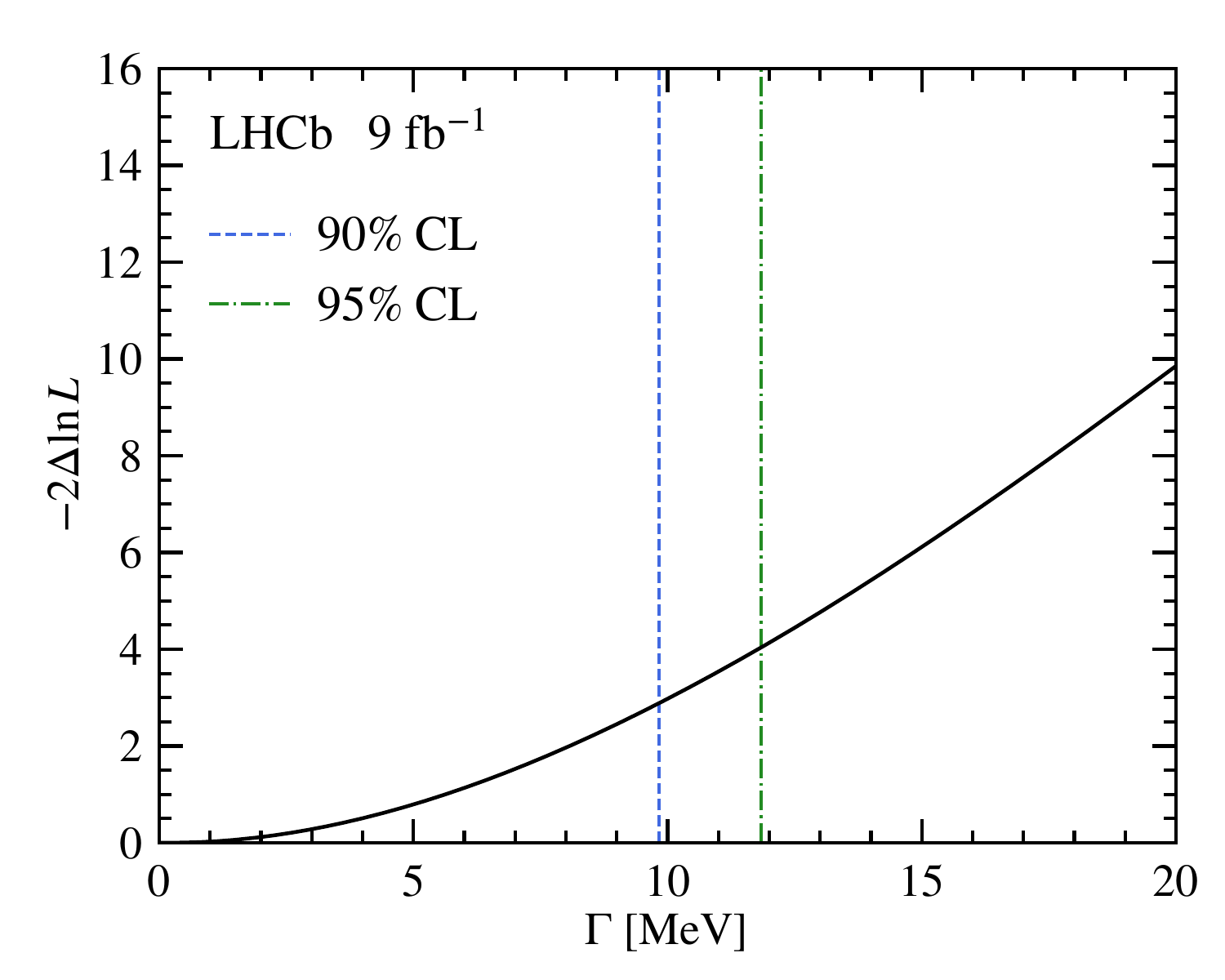}
  \caption{
   Likelihood profile of the natural width $\Gamma$ for the new state from the most conservative envelope.
   The dashed blue and dash-dotted green lines indicate the 90\% and 95\% Bayesian upper limits, respectively, derived from the posterior distribution constructed with a uniform prior.
  }
  \label{fig:GammaScan}
\end{figure}

%% ======= Fit Results & Significance
The simultaneous fit yields $343^{\,+\,44}_{\,-\,41}$ signal counts. 
The mass of the resonance is measured to be $5698.9\mevcc$,
with a $1.5\mevcc$ statistical and $0.6\mevcc$ systematic uncertainty,
where the latter accounts for fit model dependencies, track momentum calibration and external inputs on the $\Bs$ and $\piz$ masses taken from Ref.~\cite{PDG2026}. 
The natural width is found to be consistent with zero, as illustrated by the likelihood profile in Fig.~\ref{fig:GammaScan}.
Profile-likelihood scans are performed across all considered variations of the signal and background models, from which the most conservative envelope is chosen.
Using a Bayesian approach with a uniform prior~\cite{DAgostini:1999gfj} imposed over the range of 0 to $20\mev$, this is translated into an upper limit on the natural width of $\Gamma < 9.8~(11.8)\mev$ at the $90\%$ ($95\%$) confidence level~(CL). 
The local (global) signal significance, evaluated using pseudoexperiments over a mass range of $[5620,5860]\mevcc$ and cross-checked using Wilks' theorem~\cite{Wilks:1938dza}, exceeds 8 (7) standard deviations,
thereby establishing the  structure as a new excited \Bs state.

%%%%%%%%%%%%%%%%%%%%%%%%%%%%%%%%
%%%%       Conclusion
%%%%%%%%%%%%%%%%%%%%%%%%%%%%%%%%

Given its decay into two pseudoscalar mesons,
the state is most naturally identified as the long-awaited lowest-lying $0^+$ scalar $\Bs$ meson, the beauty counterpart of $\Dsstz$, designated as $\Bsstznum$ hereafter.
Notably, its mass is significantly below the conventional quark-potential model predictions~\cite{Ebert:2009ua,Sun:2014wea,Godfrey:2016nwn,Lu:2016bbk},
analogous to the mass deficit observed for the $\Dsstz$ state.
Consequently, the discovery of $\Bsstznum$ marks the first observation of a nonconventional single-beauty hadron.
The properties of $\Bsstznum$ in comparison with those of $\Dsstz$ provide a direct experimental validation of heavy-quark flavour symmetry in the nonconventional sector~\cite{Fu:2021wde}.

The observed structure is also compatible with the $1^+$ axial-vector state, the beauty counterpart of $\Dsstone$, decaying into \mbox{$\Bsst(\to \Bs \gamma) \piz$} with the photon unreconstructed, due to the small energy release in the $\Bsst \to \Bs \gamma$ transition.
Under this hypothesis, 
the state would have a mass of $5748.3\pm1.5\stat\pm 0.6\syst$,
where the latter uncertainty accounts for the current precision on the mass difference between the $\Bsst$ and $\Bs$ mesons~\cite{CMS:2025byz}.
The upper limit on its natural width would be consistent with the value obtained under the $0^+$  hypothesis.
The structure could also be a combination of both the $0^+$ and $1^+$ states~\cite{BaBar:2006eep} if their mass difference matches that between $\Bsst$ and $\Bs$ mesons, as observed in the charm sector $[M(\Dsstone)-M(\Dsstz)] - [M(\Dss)-M(\Ds)] = -2.7\pm0.7\mevcc$~\cite{PDG2026, LHCb-PAPER-2026-019, BESIII:2025ksu}.
As the scalar and axial-vector states form a \Bs doublet,
either interpretation represents a consistent departure from the conventional quark-potential model predictions.

Under the $0^+$ assignment, the mass splitting of $332.0\pm1.5\stat\pm0.6\syst\mevcc$ between $\Bsstznum$ and the ground-state $\Bs$ is close to the $349.2\pm0.7\mevcc$ splitting for its charm-sector counterpart~\cite{PDG2026}. 
This approximate flavour independence supports the universal nature of chiral dynamics at the heavy-quark scale, aligned with expectations from the chiral doubling mechanism~\cite{Nowak:1992um,Bardeen:1993ae,Bardeen:2003kt,Nowak:2003ra}. The observation of a light and narrow excited \Bs state is also in good agreement with predictions from models featuring a significant hadronic molecular component~\cite{Cleven:2010aw,Fu:2021wde,Zhou:2020moj,Altenbuchinger:2013vwa,Zhang:2024usz,Ni:2023lvx,Guo:2021rjv,Albaladejo:2016ztm,Ortega:2016pgg}, which are further supported by lattice QCD calculations~\cite{Gregory:2010gm,Lang:2015hza,Wurtz:2015mqa,Hudspith:2023loy}
as well as frameworks incorporating chiral symmetry~\cite{Colangelo:2012xi,Wang:2015mxa,Alhakami:2020vil,Gandhi:2022nnk,Colangelo:2005gb,Badalian:2007yr}.
Furthermore, the $\Bsstznum$ observation provides the first experimental determination of the lowest-lying $0^+$ pole for the \Bs meson, enabling a data-driven description of the heavy-to-light transition form factors across the full kinematic range~\cite{Melic:2025uha, Bharucha:2010im, Biswas:2025drz}.

The future identification of the axial-vector $1^+$ state will be essential to establish the $(0^+,1^+)$ \Bs doublet. Together with the $0^+$ state, it would not only consolidate the relevant theoretical models, but also provide a precise handle to quantify the degree of heavy-quark spin-symmetry breaking in the exotic domain.
Future explorations of the $\Bs\gamma$ invariant-mass spectrum will serve as a definitive probe towards an unambiguous mapping of the doublet.
The $0^+$ scalar state should produce a peak at $5649.5\mevcc$ through the $\Bsst\gamma$ decay, with the photon from the $\decay{\Bsst}{\Bs\gamma}$ decay left unreconstructed.
Meanwhile, the $1^+$ axial-vector state should manifest itself in a form of two distinct peaks at $5748.3$ and $5698.9\mevcc$ corresponding to its decays into $\Bs\gamma$ and $\Bsst\gamma$ final states, respectively. The relative yields of the two peaks would be sensitive to the nature of the state~\cite{Bardeen:2003kt,Fu:2021wde}.
All three predicted peak positions carry an uncertainty of $1.6\mevcc$, dominated by the precision of the present $\Bsstznum$ mass measurement.
Within the chiral doubling framework, the mass splitting between chiral partners measured in the $\Ds$ and $\Bs$ systems can be used to predict the masses of other heavy-light systems, such as the $D$, $B$ mesons, as well as the doubly heavy \Xicc and \Omegacc baryons since the double-heavy diquark in an antitriplet colour state can be viewed, from the QCD perspective~\cite{Savage:1990di}, as a heavy antiquark. Based on the mass splitting measured in this analysis, 
the masses for the yet-undiscovered lowest-lying $0^+$ and $1^+$ states in the nonstrange $B$ sector are expected to be $5611.5\pm1.6\mevcc$ and $5656.8\pm1.6\mevcc$, respectively.
Both states are anticipated to be broad owing to kinematically favoured and isospin-allowed $\B^{(*)}\pi$ decays. 
Similarly, broad partners of the ground-state \Xiccp and \Xiccpp baryons~\cite{LHCb-PAPER-2017-018,LHCb-PAPER-2026-009} are expected to have masses between 3952 and 3969\mevcc and to decay predominantly through the \Xicc\pion modes.
Analogous to the \Ds and \Bs systems, a distinctly narrow partner for the newly discovered \Omegacc baryon~\cite{LHCb-PAPER-2026-022} is predicted with a mass below the $\Xicc K$ threshold featuring isospin-violating decay into the $\Omegacc\piz$ final state. 

In summary,
a new resonance is observed in the $\Bs \piz$ invariant-mass spectrum using the $pp$ collision data collected by the \lhcb experiment between 2011 and 2018,
corresponding to an integrated luminosity of $9\invfb$.
Its mass is measured to be
\begin{equation*}
    M = 5698.9\pm1.5\stat\pm 0.6\syst \mevcc,
\end{equation*}
where the first uncertainty is statistical and the second is systematic,
and its width is found to be
\begin{equation*}
    \Gamma < 9.8~(11.8)\mev \, \text{at $90\%$ ($95\%$) CL} .
\end{equation*}
The discovery of an unexpectedly light and narrow beauty-strange state reveals a significant mass deficit against quark-potential model predictions, establishing a consistent anomaly across both charm and beauty sectors.
With a multifold increase of the dataset size compared to this analysis, 
the data collected by the upgraded \lhcb experiment~\cite{LHCb-TDR-012} since 2022 will be key to resolving the $(0^+,1^+)$ $\Bs$ doublet and searching for analogous states in other heavy-hadron systems,
thereby offering deeper insights into the chiral dynamics and colour confinement mechanism governing hadron spectroscopy.

\clearpage
\section*{Methods}
\label{sec:methods}

%%%%%%%%%%%%%%%%%%%%%%%%%%%%%%%%
%%%%        Detector
%%%%%%%%%%%%%%%%%%%%%%%%%%%%%%%%
\subsection*{Experimental setup} 
The \lhcb detector~\cite{LHCb-DP-2008-001,LHCb-DP-2014-002} is a single-arm forward spectrometer covering the pseudo-rapidity range $2<\eta<5$, designed for the study of particles containing $\bquark$ or $\cquark$ quarks.
 The pseudorapidity $\eta$ is defined as $-\ln(\tan(\theta / 2))$, where $\theta$ is the polar angle of the track relative to the proton beam line.
Charged-particle trajectories are measured by a tracking system comprising a vertex detector surrounding the interaction region, a large-area silicon-strip detector located upstream of a dipole magnet with a bending power of about $4{\mathrm{\,T\,m}}$, whose polarity is regularly reversed so that data with both magnet polarities contribute to the analysis, and three downstream stations of silicon-strip and straw-tube detectors.
The momentum of charged particles is measured with a relative uncertainty that ranges from about 0.5\% at low momentum to 1.0\% at 200\gevc.
Charged hadrons are identified by two ring-imaging Cherenkov detectors that separate pions, kaons and protons over the relevant momentum range.
Photons, electrons and hadrons are then distinguished by a calorimeter system comprising scintillating-pad and pre-shower detectors, an electromagnetic calorimeter~(ECAL) and a hadronic calorimeter.
The ECAL employs the Shashlik technology, \ie a sampling structure of alternating scintillator and lead layers readout by plastic wavelength-shifting fibres,
with a design energy resolution of $\sigma_{E}/E=10\%/\sqrt{E}\oplus1\%$~($E$ in\gev)~\cite{LHCb-TDR-002}.
The ECAL energy scale undergoes an initial calibration via the energy flow method,
followed by a fine calibration based on the $\decay{\piz}{ \gamma \gamma}$ decay~\cite{Belyaev:2014zga}.
Muons are identified by a system composed of alternating layers of iron and multiwire proportional chambers.

The online event selection is performed by a two-level trigger system~\cite{LHCb-DP-2012-004,LHCb-DP-2019-001}. A first hardware stage uses information from the calorimeter and muon systems to select high-transverse-energy clusters or high-\pt muons, where \pt is the component of the momentum transverse to the beam. 
A subsequent software stage performs a full event reconstruction and selects heavy-flavour decays using algorithms that exploit the large impact parameters and displaced vertices characteristic of $\bquark$-hadron decays. The hadronic decays $\decay{\Bs}{ \Dsm\pip}$ and $\decay{\Bs}{ \Dsm\pip\pip\pim}$ are accepted if a multitrack displaced vertex is consistent with the decay of a $\bquark$ hadron~\cite{BBDT,LHCb-PROC-2015-018}, while for the $\decay{\Bs}{\jpsi \Pphi}$ decay mode the trigger requires a pair of oppositely charged muons forming a good vertex with an invariant mass compatible with the $\jpsi$ meson.

%%%%%%%%%%%%%%%%%%%%%%%%%%%%%%%%
%%%%       Simulation
%%%%%%%%%%%%%%%%%%%%%%%%%%%%%%%%
\subsection*{Simulated samples}
Simulation is required to model the signal behavior, as well as the effects of detector acceptance, resolution, and to optimise the selection requirements. In the simulation, $\proton\proton$ collisions are generated using \pythia~\cite{Sjostrand:2007gs} with a specific \lhcb configuration~\cite{LHCb-PROC-2010-056}. 
The decays of unstable particles are modelled using \evtgen~\cite{Lange:2001uf}, and final-state radiation is generated using \photos~\cite{davidson2015photos}. 
The interaction of the generated particles with the detector, and its response,
are implemented using the \geant
toolkit~\cite{Allison:2006ve, *Agostinelli:2002hh} as described in
Ref.~\cite{LHCb-PROC-2011-006}. 
The simulated sample is corrected to match the track reconstruction efficiency~\cite{LHCb-DP-2013-002}, particle kinematics, event activity, and particle-identification~(PID) response~\cite{LHCb-DP-2013-001} in data.
In simulation, the PID responses are sampled from calibration samples of $\decay{\Dstarp}{ \Dz (\to \Km \pip) \pip}$ and $\decay{\KS}{ \pip \pim}$ decays for charged pions and kaons,
and $\decay{\jpsi}{ \mumu}$ for muons.
The $\Bs\piz$ mass resolutions are determined from simulated $\Bsstz \to \Bs\piz$ decays, initially generated with a $\Bsstz$ mass of $5707\mevcc$.
The impact of the mass difference relative to the measured mass of $\Bsstznum$ is confirmed to be negligible using {\sc{RapidSim}}~\cite{Cowan:2016tnm}.

%%%%%%%%%%%%%%%%%%%%%%%%%%%%%%%%
%%%%       Selection
%%%%%%%%%%%%%%%%%%%%%%%%%%%%%%%%
\subsection*{Candidate selection} 
In each of the three $\Bs$ decay modes, charged particles are required to have good track quality and PID responses compatible with the kaon, pion or muon hypotheses. Intermediate $\Dsm$, $\jpsi$ and $\Pphi$ candidates are reconstructed within windows around their known masses~\cite{PDG2026} and must form well-resolved vertices significantly displaced from the $pp$ interaction point. Background candidates from misidentified charmed-hadron decays, such as those of $\Dm$, $\Dz$ and $\Lc$, are suppressed with vetoes that combine invariant-mass requirements under alternative particle hypotheses with PID selections, while retaining high efficiency for genuine $\Bs$ decays. 

After these preselection requirements, BDT classifiers~\cite{AdaBoost,XGBoosting} are applied separately in each $\Bs$ decay mode to enhance the purity of the $\Bs$ samples. The input variables describe the quality and displacement of the $\Bs$ vertex, the pointing of the $\Bs$ momentum towards the $pp$ interaction point, and kinematic, impact-parameter and PID observables. Signal training samples are taken from the simulated $\Bs$ decays described above, while background samples are drawn from high-mass sidebands of the $\Bs$ invariant-mass distribution. The selection thresholds on the BDT outputs are chosen to optimise the $\Bs$ signal yields, using simulated signal decays and sideband data to estimate the signal efficiency and background yield, respectively.

To form $\Bs\piz$ candidates, selected $\Bs$ candidates are combined with $\piz$ candidates, which are reconstructed from a pair of photons identified as isolated clusters in ECAL.  
The subsequent $\Bs\piz$ selection is optimised separately for each of the three $\Bs$ decay modes to account for their differing background levels and mass resolutions.
A kinematic fit constraining the $\Bs$ and $\piz$ decay products to a common vertex and the reconstructed $\Dsm$, $\jpsi$, $\Bs$ and $\piz$ masses to their known values is performed.
Only candidates with good fit quality and a fitted $\Bs$ mass compatible with the known mass~\cite{PDG2026} are retained. 
In addition, the $\piz$ candidate is required to have an invariant mass close to its known mass~\cite{PDG2026} and sufficient \pt, and the photons from the $\piz$ candidate must satisfy loose shower-shape and cluster-quality criteria to suppress poorly reconstructed neutral pions. 
A second set of BDT classifiers is trained at the $\Bs\piz$ level, using variables related to the $\Bs\piz$ and $\piz$ candidate kinematics and the properties of the photons forming the $\piz$ candidate. 
Signal samples are taken from simulated $\decay{\Bsstznum}{ \Bs\piz}$ decays corrected to reproduce the kinematic and event-activity distributions observed in data, while background samples are obtained from $\piz$ mass sidebands in data, which provide a large sample of combinatorial $\Bs\piz$ candidates.

The selection criteria on the classifiers are optimised by maximising the figure of merit described in Ref.~\cite{Punzi:2003bu}, with the signal efficiency taken from simulated $\decay{\Bsstznum}{ \Bs\piz}$ decays. 
The expected background in the search window is estimated from a mixed-event $\Bs\piz$ template constructed by combining $\Bs$ and $\piz$ candidates from different events to model uncorrelated combinatorial background and normalised to data in the control region $[5940,6400]\mevcc$.
In events containing multiple $\Bs\piz$ combinations, typically when several $\piz$ candidates are associated with the same $\Bs$ candidate, only one entry per event with the reconstructed $\piz$ mass closest to the known value~\cite{PDG2026} is retained.

\begin{figure}[t]
  \centering
  \begin{minipage}{0.32\textwidth}
    \centering
    \textbf{\phantom{00}$\decay{\Bs}{ \Dsm\pip}$}\\[0.2cm]
    \includegraphics[width=\textwidth]{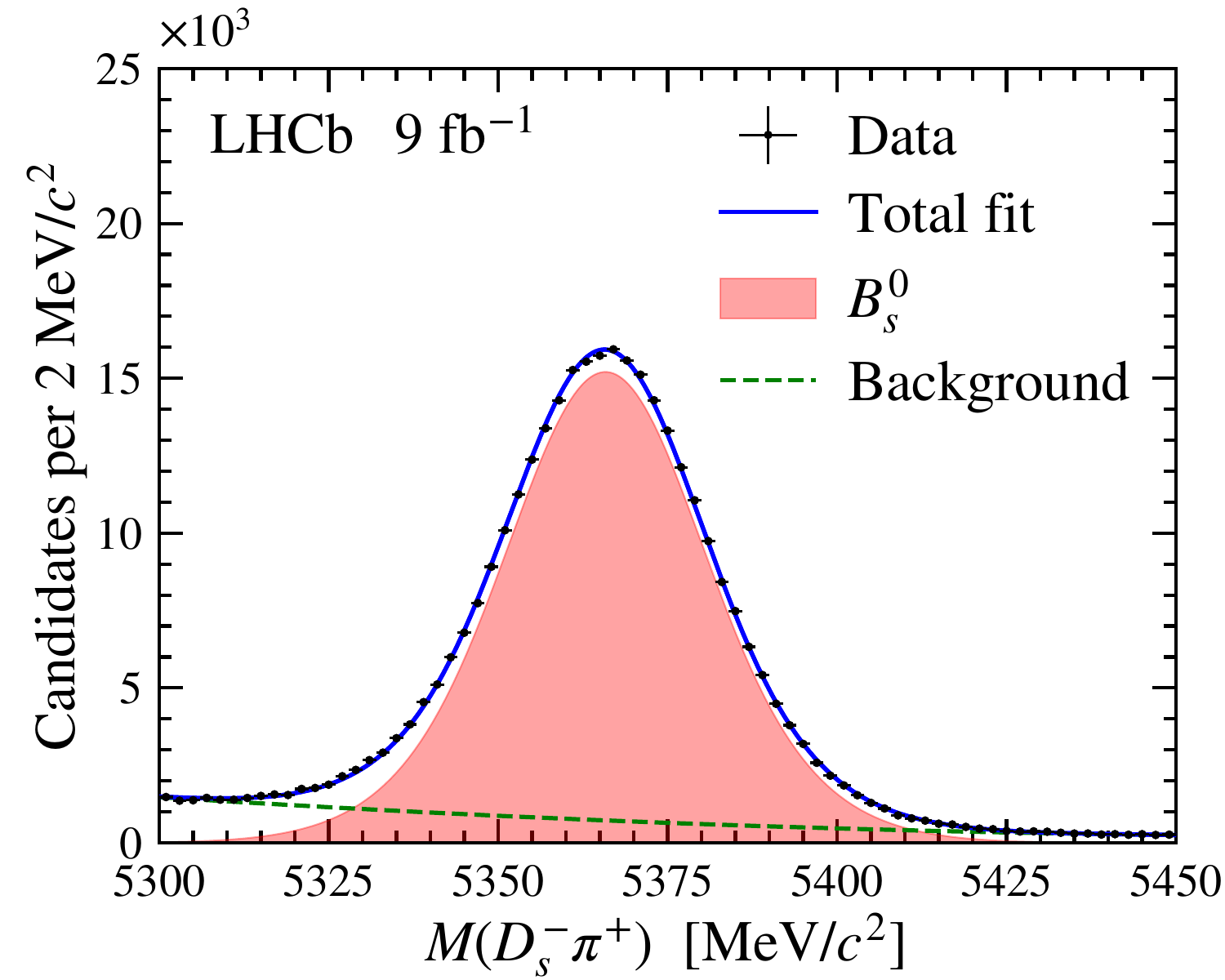}
  \end{minipage}
  \hfill
  \begin{minipage}{0.32\textwidth}
    \centering
    \textbf{\phantom{00}$\decay{\Bs}{ \jpsi\phi}$}\\[0.2cm]
    \includegraphics[width=\textwidth]{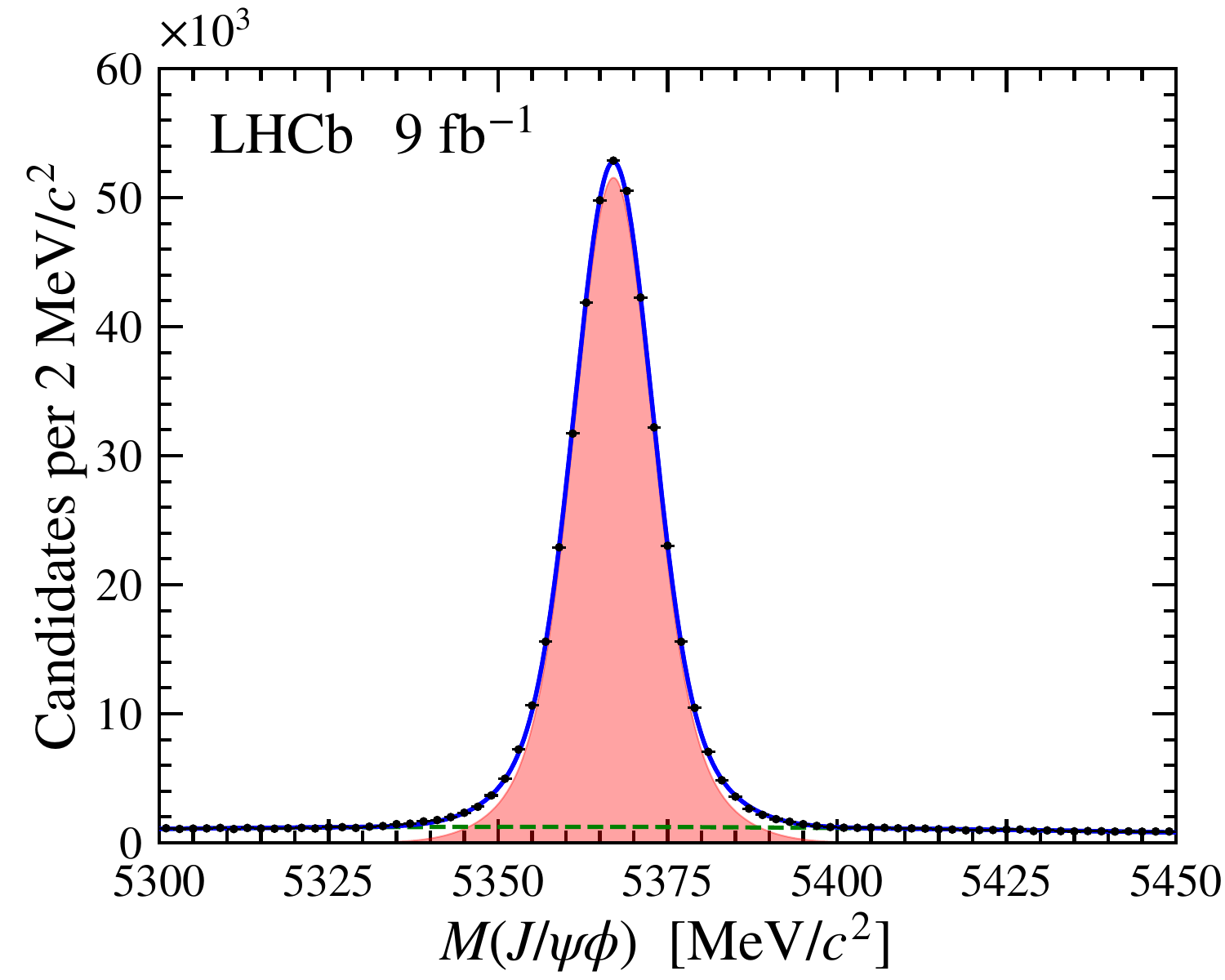}
  \end{minipage}
  \hfill
  \begin{minipage}{0.32\textwidth}
    \centering
    \textbf{\phantom{00}$\decay{\Bs}{ \Dsm\pip\pip\pim}$}\\[0.2cm]
    \includegraphics[width=\textwidth]{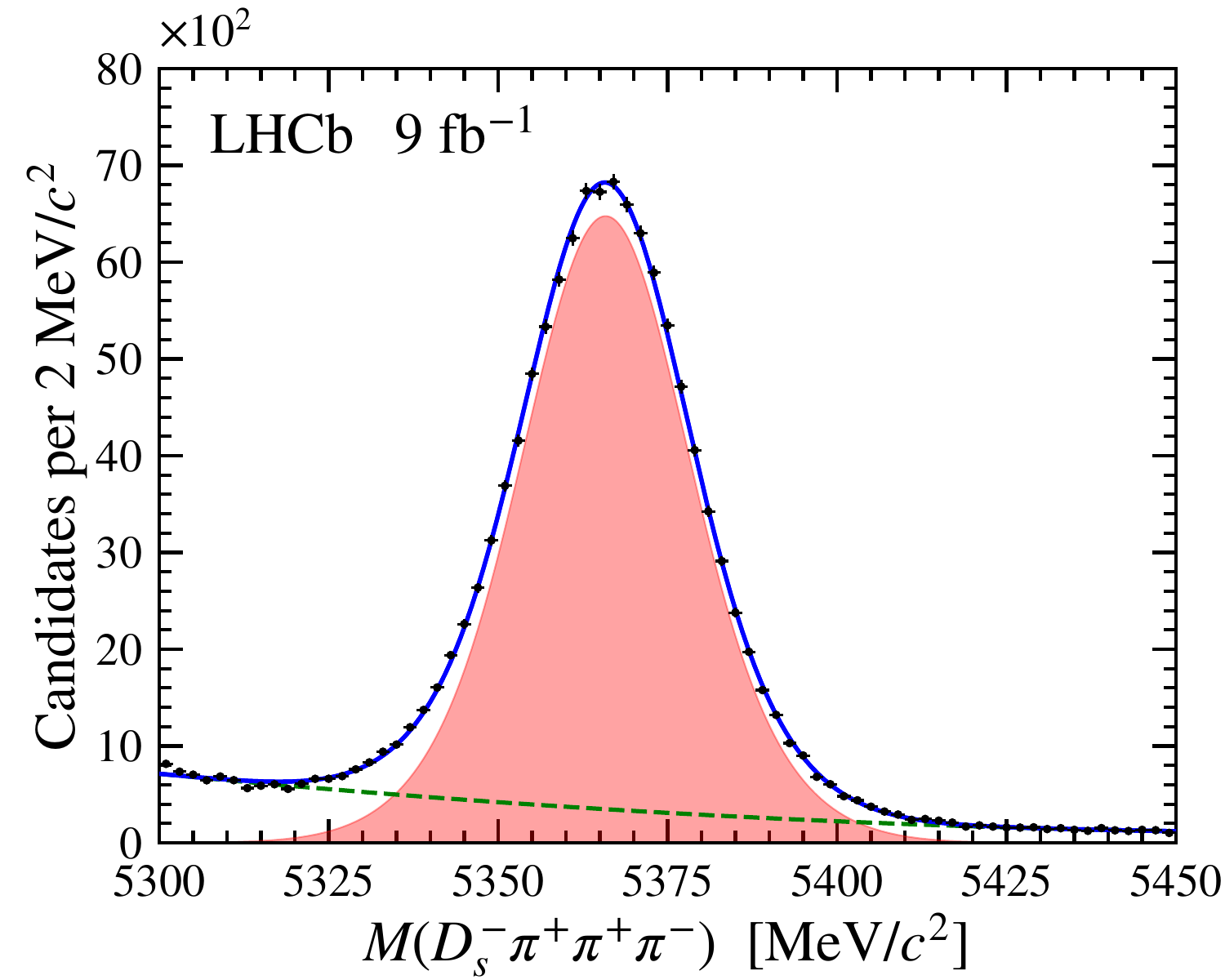}
  \end{minipage}

  \begin{minipage}{0.32\textwidth}
    \centering
    \includegraphics[width=\textwidth]{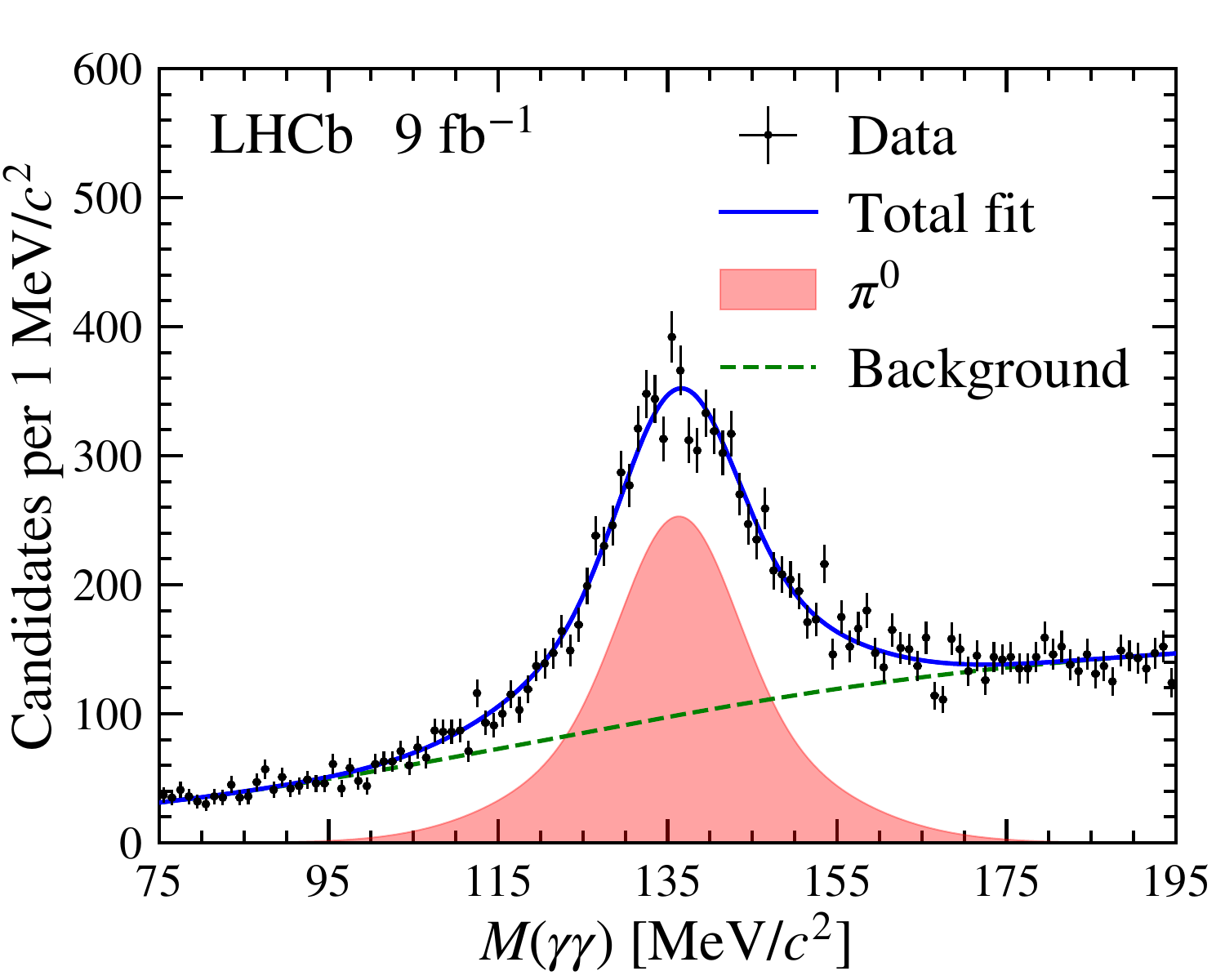}
  \end{minipage}
  \hfill
  \begin{minipage}{0.32\textwidth}
    \centering
    \includegraphics[width=\textwidth]{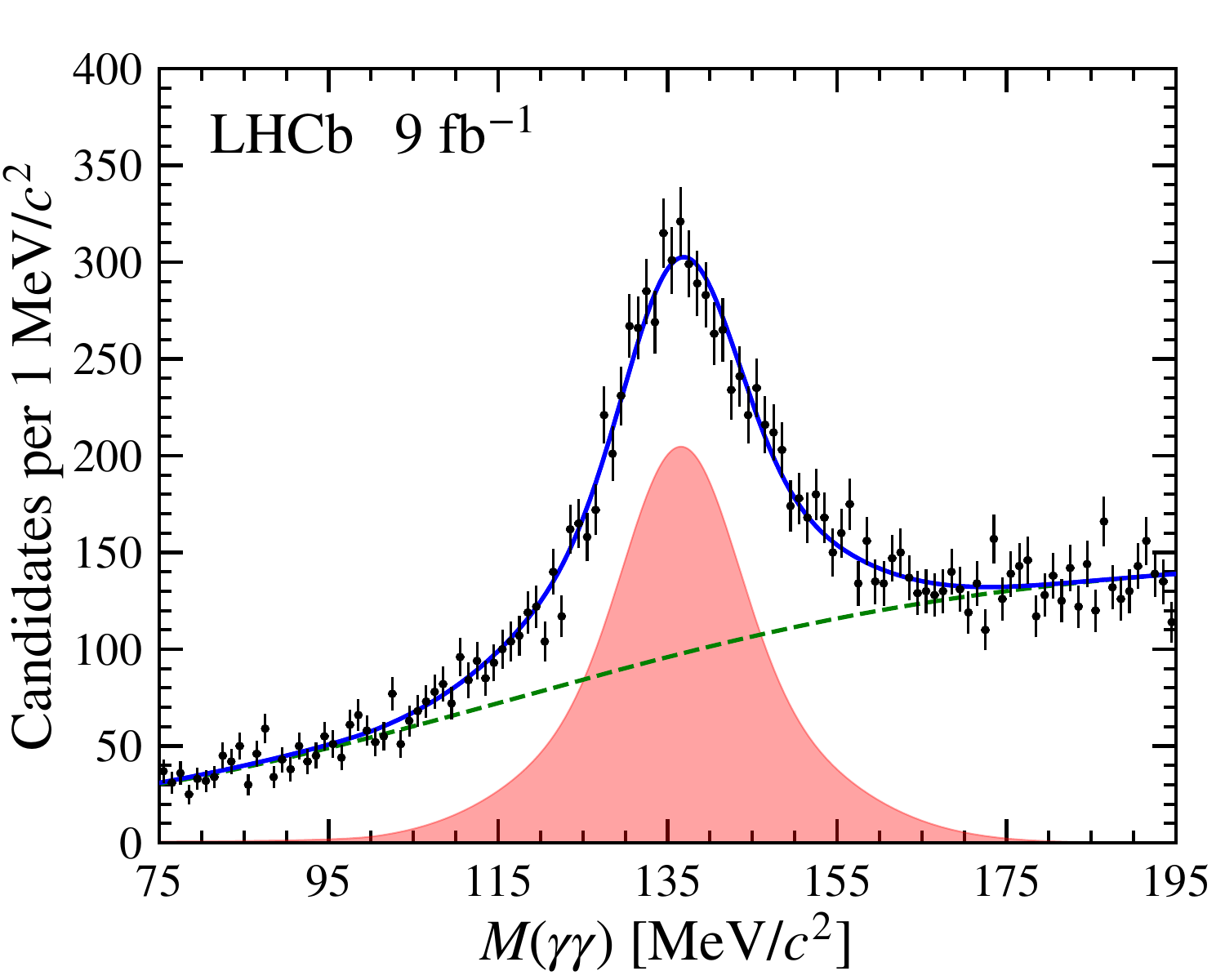}
  \end{minipage}
  \hfill
  \begin{minipage}{0.32\textwidth}
    \centering
    \includegraphics[width=\textwidth]{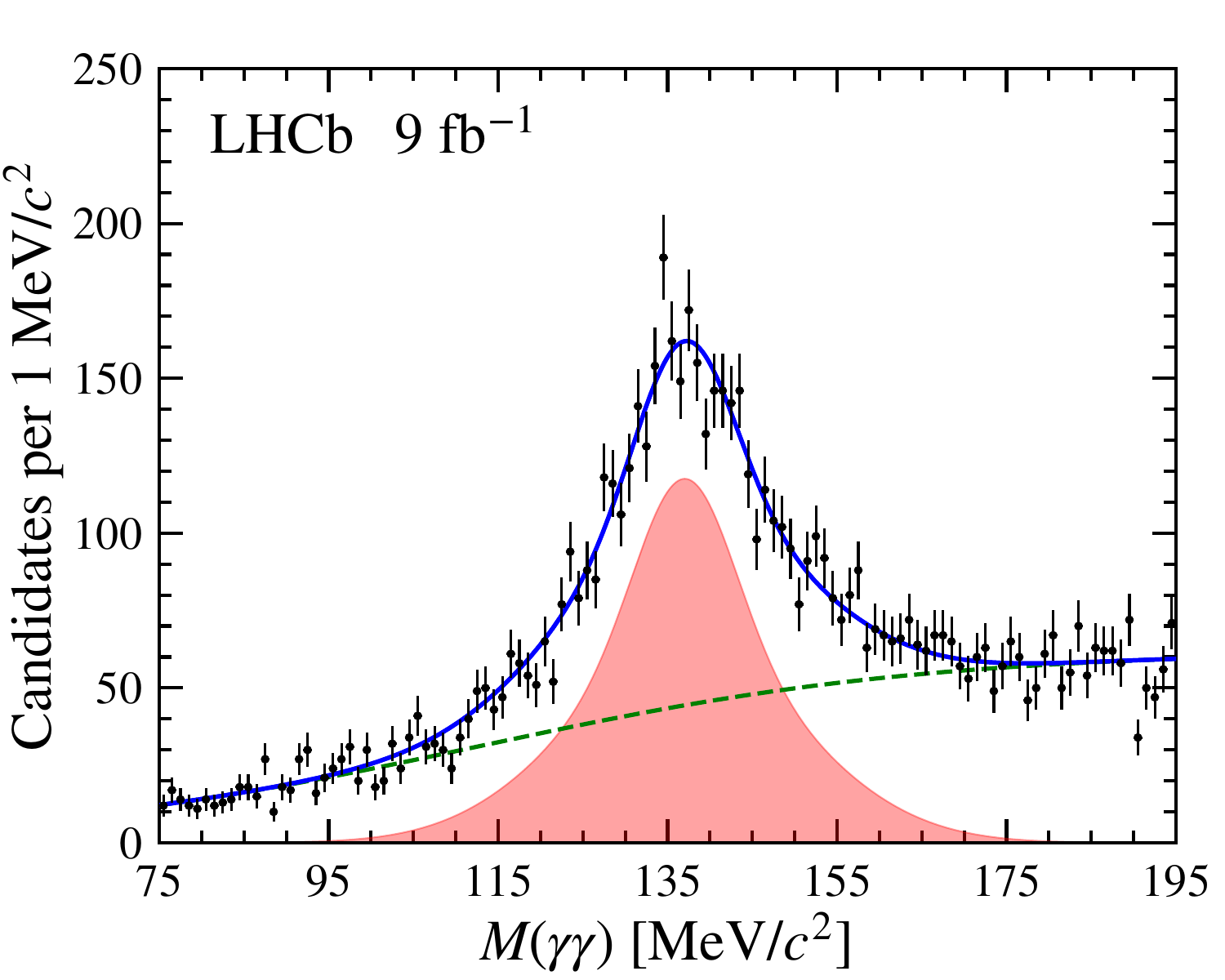}
  \end{minipage}

  \begin{minipage}{0.32\textwidth}
    \centering
    \includegraphics[width=\textwidth]{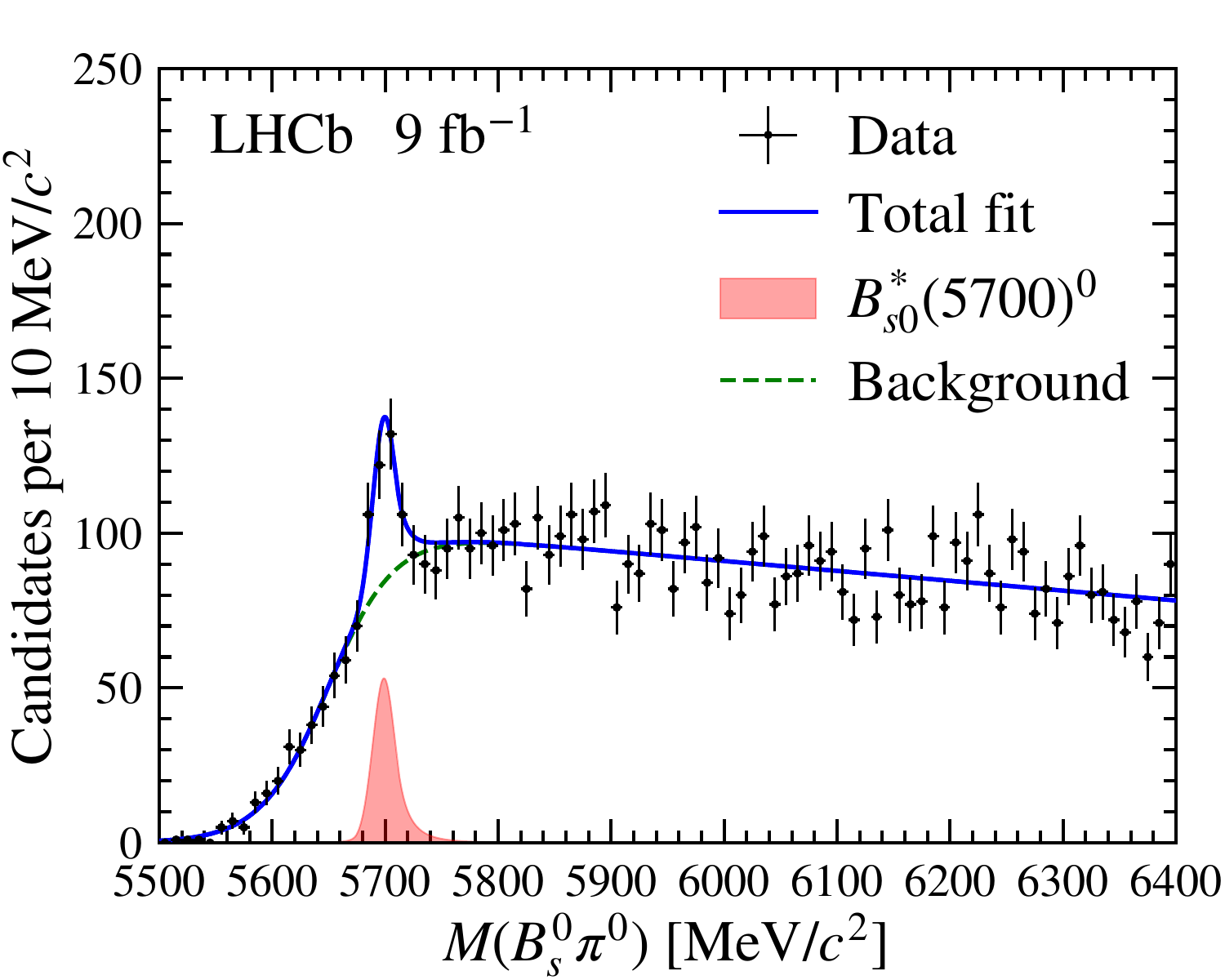}
  \end{minipage}
  \hfill
  \begin{minipage}{0.32\textwidth}
    \centering
    \includegraphics[width=\textwidth]{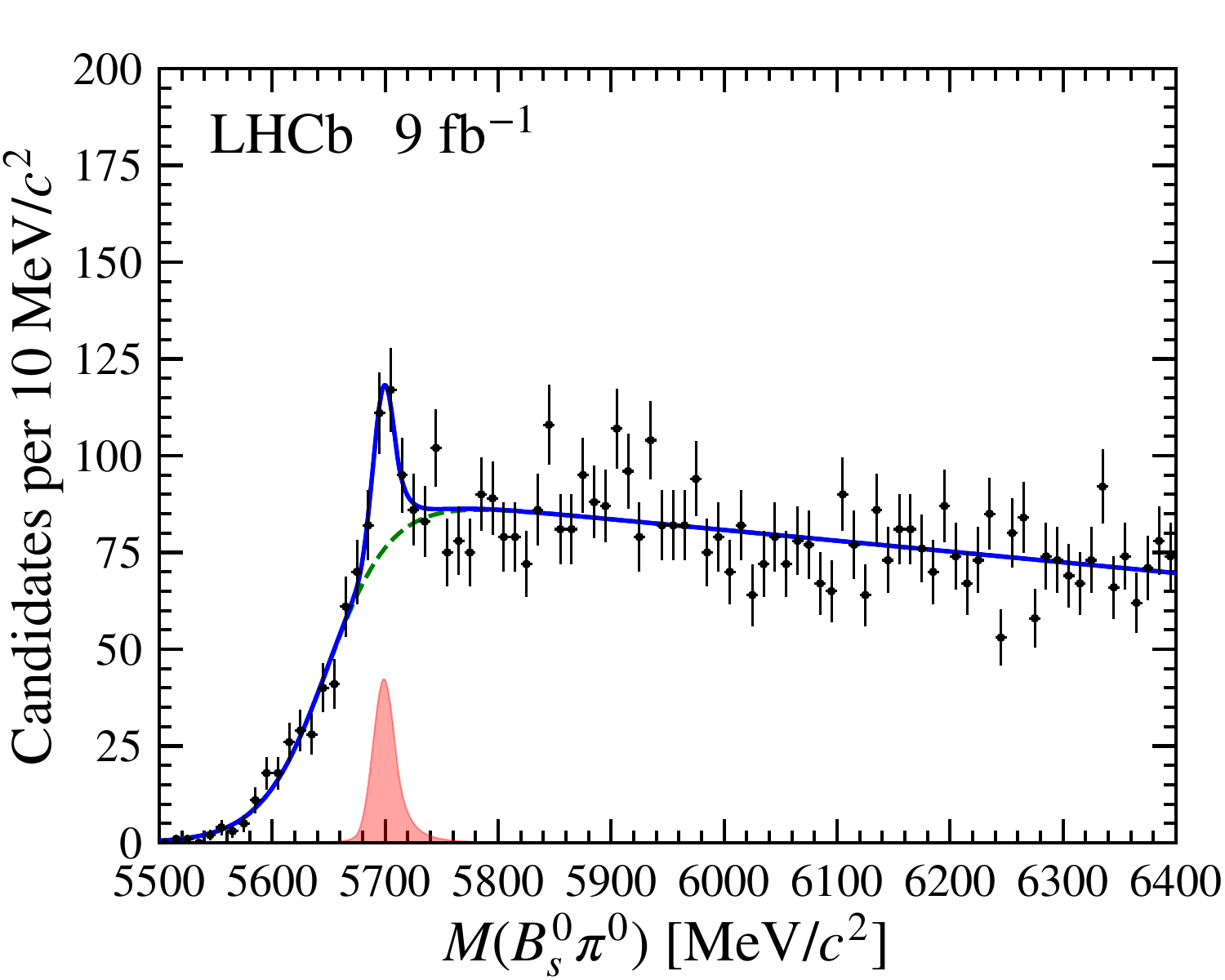}
  \end{minipage}
  \hfill
  \begin{minipage}{0.32\textwidth}
    \centering
    \includegraphics[width=\textwidth]{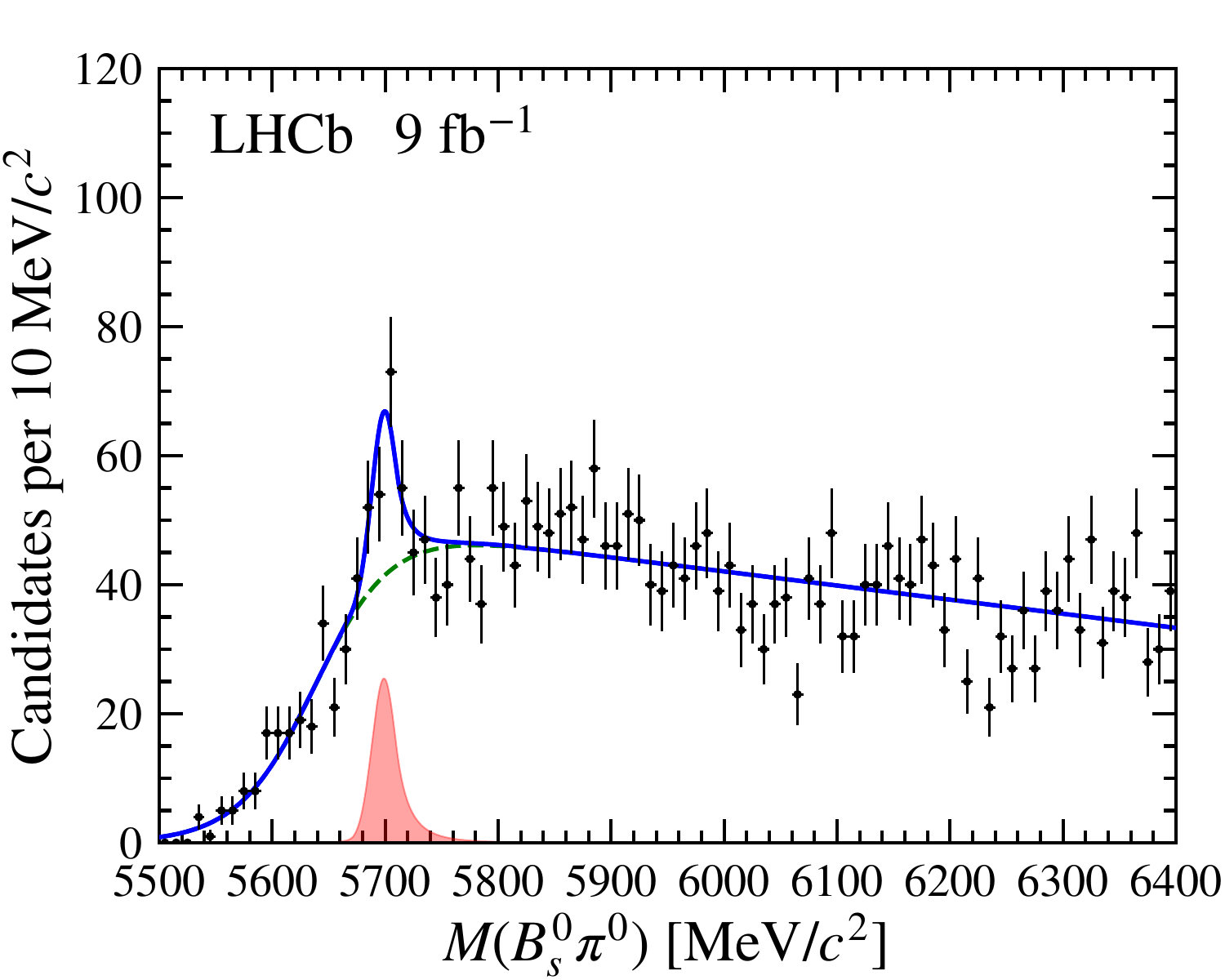}
  \end{minipage}

  \caption{ 
    Invariant-mass distributions of  reconstructed (top) \Bs , (middle) \piz, and (bottom) \Bs\piz candidates for the three \Bs decay modes: (left) $\decay{\Bs}{ \Dsm\pip}$, (middle) $\decay{\Bs}{ \jpsi\phi}$, and (right) $\decay{\Bs}{ \Dsm\pip\pip\pim}$. Only \piz candidates from selected $\Bs\pi^0$ combinations with invariant mass in the $[5500, 6400]\mevcc$ range are shown. Fit results to the invariant-mass distributions are overlaid.
    }
  \label{fig:massfit_3modes}
\end{figure}

%%%%%%%%%%%%%%%%%%%%%%%%%%%%%%%%
%%%%       Signal yields
%%%%%%%%%%%%%%%%%%%%%%%%%%%%%%%%
\subsection*{Signal yields} 
The invariant-mass distributions of the selected $\Bs$ candidates for the three decay modes are shown in the top row of Fig.~\ref{fig:massfit_3modes}.
To determine the signal yields, unbinned maximum-likelihood fits are performed using a combination of a Gaussian kernel with power-law tails for the signal and polynomial functions to model the combinatorial background.
The signal yields of $\decay{\Bs}{\Dsm\pip}$, $\decay{\Bs}{\jpsi\phi}$ and $\decay{\Bs}{\Dsm\pip\pip\pim}$ decay modes are measured to be 
\mbox{$(2.96\pm0.02)\times 10^5$},
\mbox{$(4.16\pm0.01)\times 10^5$}, and
\mbox{$(1.06\pm0.01)\times 10^5$}, respectively.

After combining the $\Bs$ and $\piz$ candidates, and applying the optimised selection criteria,
the $\piz$ invariant-mass distributions for selected $\Bs\piz$ candidates falling in the $\Bs\piz$ mass fit region of $[5500,6400]\mevcc$ are shown in the middle row of Fig.~\ref{fig:massfit_3modes}.
A clear $\piz$ mass peak is observed, modelled by a Gaussian kernel with power-law tails on top of a combinatorial background described by a sigmoid function modulated by a polynomial function.

The $\Bs\piz$ invariant-mass distributions for each individual decay mode are displayed in the bottom row of Fig.~\ref{fig:massfit_3modes},
with the simultaneous fit projections as described in the main text overlaid.
A narrow resonant peak is consistently observed at an identical position across all three independent decay modes,
yielding $\Bsstznum$ signal counts of
$148^{\, + \, 29}_{\, - \, 26}$ for $\decay{\Bs}{ \Dsm\pip}$,
$116^{\, + \, 26}_{\, - \, 25}$ for $\decay{\Bs}{ \jpsi\phi}$,
and $79^{\, + \, 21}_{\, - \, 19}$ for $\decay{\Bs}{ \Dsm\pip\pip\pim}$.
Extensive checks using event-mixing techniques and sideband control samples exhibit no comparable peaking structure,
confirming that the peak does not originate from kinematic reflections or combinatorial artifacts.
Furthermore, the signals are consistent across different data-taking periods and \lhcb dipole magnet polarities,
with all fit results aligning with expectations.

\subsection*{Resolution scale factor}
     
The decays of the $\Sigmac(2455)^{+}$, $\Sigmac(2520)^{+}$ and \Xicp baryons to the common  \Lc \piz final state are studied to account for potential discrepancies in the detector resolution between data and simulation. These control modes have a similar energy release as the $\decay{\Bsstznum}{  \Bs \piz}$ signal, making them well suited for assessing potential mismodelling of the mass resolution in simulation.
This study is performed using a dataset of $5.4\invfb$, representative of the conditions across the full data sample, as it provides sufficient statistical power and yields consistent resolution corrections.
The $\Lc$ candidates are reconstructed via the $\proton\Km\pip$ final state, and combined with $\decay{\piz}{ \gamma\gamma}$ candidates selected using similar criteria as for the $\Bs\piz$ signal. A BDT classifier is employed to suppress combinatorial background, optimised using the figure of merit of Ref.~\cite{Punzi:2003bu}.

\begin{figure}[!t]
\centering
\includegraphics[width=0.65\textwidth]{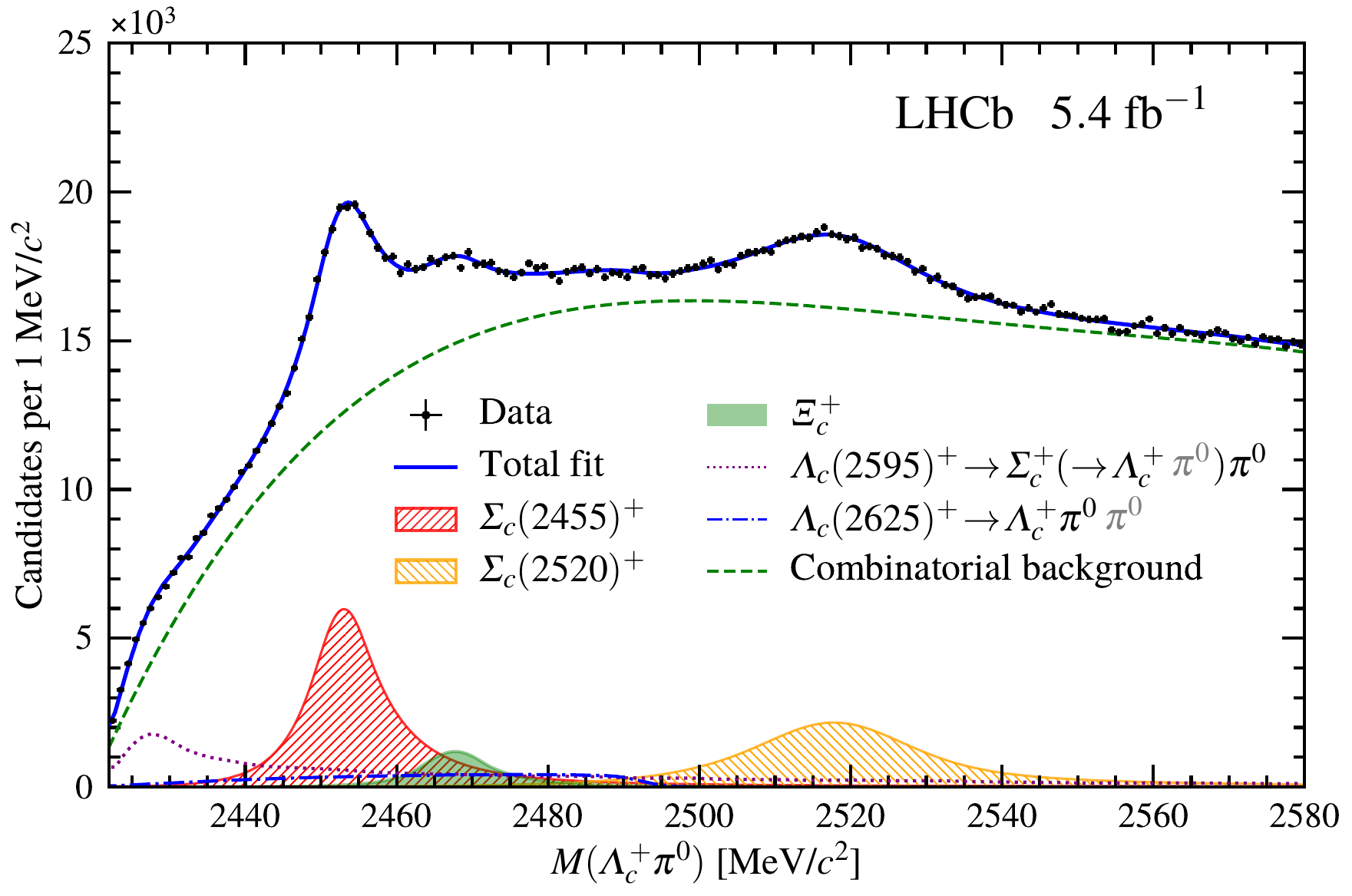}
\caption{
Invariant-mass distribution of $\Lc\piz$ candidates used for resolution calibration, with the fit results superimposed. 
The legend specifies the components, 
where the grey $\piz$ symbol indicates the unreconstructed pion, 
while hashed and dashed histograms denote the signal and partially reconstructed components, respectively.
}
\label{fig:massfit_Lcpi0}
\end{figure}

An extended unbinned maximum-likelihood fit is performed to the $\Lc\piz$ mass spectrum, as shown in Fig.~\ref{fig:massfit_Lcpi0}. The $\Sigmac(2455)^{+}$ and $\Sigmac(2520)^{+}$ signals are modelled by relativistic Breit--Wigner functions convolved with the simulated resolution functions. 
The fit model also accounts for the $\decay{\Xicp}{\Lc \piz}$ peak, modelled by a simulated resolution function, and feed-down contributions from other decays, such as $\decay{\Lz_{\cquark}(2595)^+}{ \Sigmac(2455)^+(\to \Lc\piz)\piz}$ and $\decay{\Lz_{\cquark}(2625)^+}{ \Lc\piz\piz}$ where one $\piz$ meson is not reconstructed. 
The background is described by a fourth-order polynomial function. 
The mass resolution in data is parametrised as the corresponding resolution in simulation scaled by a factor which is left as a free parameter.
The scale factor is determined to be $1.03 \pm 0.05$, which is subsequently applied to the $\Bsstznum$ signal model to ensure an accurate estimation of the mass and the upper limit on its natural width.

\subsection*{Systematic uncertainty on the mass measurement}
Several sources of systematic uncertainty on the mass measurement are considered.
The uncertainty due to the potential imperfect signal model is assessed by varying the parameters of the nominal relativistic Breit--Wigner lineshape and the tail parameters of the resolution function~\cite{Skwarnicki:1986xj}.
Specifically, the orbital angular momentum is increased from the baseline value of 0 to 1 and 2 to account for potential higher orbital components and the Blatt--Weisskopf barrier radius~\cite{Blatt:1952ije} is varied from its nominal $3\gev^{-1}\hbar c$ to $5\gev^{-1}\hbar c$ to assess the sensitivity to the effective hadronic size.
The uncertainty due to the imperfect simulation of the detector resolution, accounted for by the resolution scale factor,
is included in the baseline fit and accounted for as part of the statistical uncertainty.
For the background model, the nominal function is replaced by alternative functions with varying polynomial orders and different parameterizations of the sigmoid factor.
Furthermore, the robustness of the nominal model and its alternatives are validated against potential overparameterization through a stability scan of the upper fit boundaries.
Finally, the uncertainty from the momentum scale, calibrated using well-known resonances such as $\Dz$ and $\jpsi$ mesons~\cite{LHCb-DP-2023-003}, is propagated to the $\Bs\piz$ mass measurement.
The total uncertainty amounts to $0.6\mevcc$ on the mass measurement.

\clearpage

\section*{Acknowledgements}
%
% These Acknowledgements valid from 16/06/2026
%
\noindent We express our gratitude to our colleagues in the CERN
accelerator departments for the excellent performance of the LHC. We
thank the technical and administrative staff at the LHCb
institutes.
We acknowledge support from CERN and from the national agencies:
ARC (Australia);
CAPES, CNPq, FAPERJ and FINEP (Brazil); 
MOST and NSFC (China); 
CNRS/IN2P3 and CEA (France);  % added CEA 26/02/2026
BMFTR, DFG and MPG (Germany);
NKFIH (Hungary);              % added 16/06/2026
INFN (Italy); 
NWO (Netherlands); 
MNiSW and NCN (Poland); 
MEC/IFA (Romania); 
%MSHE (Russia); 
MICIU and AEI (Spain);
SNSF and SER (Switzerland); 
NASU (Ukraine); 
STFC (United Kingdom); 
DOE NP and NSF (USA).
%%%%%%%%%%%%%%%%%%%%%%%%%%%%%%%%%%%%%%%%%%%%%
We acknowledge the computing resources that are provided by ARDC (Australia), 
CBPF (Brazil),
CERN, 
IHEP and LZU (China),
IN2P3 (France), 
KIT and DESY (Germany), 
INFN (Italy), 
SURF (Netherlands),
Polish WLCG (Poland),
IFIN-HH (Romania), % http://dx.doi.org/10.13039/100019931,"Institutul National de Cercetare-Dezvoltare pentru Fizica si Inginerie Nucleara 'Horia Hulubei'"
%RRCKI and Yandex LLC (Russia), 
PIC (Spain), CSCS (Switzerland), 
GridPP (United Kingdom),
and NSF (USA).  % added Feb2026
%%%%%%%%%%%%%%%%%%%%%%%%%%%%%%%%%%%%%%%%%% 
We are indebted to the communities behind the multiple open-source
software packages on which we depend.
%%%%%%%%%%%%%%%%%%%%%%%%%%%%%%%%%%%%%%%%%%
Individual groups or members have received support from
% ARC and ARDC (Australia); % moved to national 16/01/2025
RTP (Australia), % added 06/03/2026
FWO Odysseus grant G0ASD25N (Belgium), % added 20/4/2026
Key Research Program of Frontier Sciences of CAS, CAS PIFI, CAS CCEPP (China); 
%Fundamental Research Funds for the Central Universities,  and Sci.\
%\& Tech.\ Program of Guangzhou (China); Removed 24/11/25
Minciencias (Colombia);
EPLANET, Marie Sk\l{}odowska-Curie Actions, ERC and NextGenerationEU (European Union);
A*MIDEX, ANR, IPhU and Labex P2IO, and R\'{e}gion Auvergne-Rh\^{o}ne-Alpes (France);
%RFBR, RSF and Yandex LLC (Russia);
Alexander-von-Humboldt Foundation (Germany);
ICSC (Italy); 
%GVA, XuntaGal, GENCAT, Inditex, InTalent and Prog.~Atracci\'on Talento, CM (Spain);
Severo Ochoa and Mar\'ia de Maeztu Units of Excellence, GVA, XuntaGal, GENCAT, InTalent-Inditex and Prog.~Atracci\'on Talento CM (Spain);
%XuntaGal --> Xunta de Galicia 
% SRC (Sweden);  % removed 27/02/2026 - end of grant
the Leverhulme Trust, the Royal Society and UKRI (United Kingdom).

\addcontentsline{toc}{section}{References}

\nocite{LHCb-PAPER-2012-030,LHCb-PAPER-2020-026}
\bibliographystyle{LHCb/LHCb}
\bibliography{main,LHCb/standard,LHCb/LHCb-PAPER,LHCb/LHCb-CONF,LHCb/LHCb-DP,LHCb/LHCb-TDR}

\ifx\mcitethebibliography\mciteundefinedmacro
\PackageError{LHCb.bst}{mciteplus.sty has not been loaded}
{This bibstyle requires the use of the mciteplus package.}\fi
\providecommand{\href}[2]{#2}
\begin{mcitethebibliography}{100}
\mciteSetBstSublistMode{n}
\mciteSetBstMaxWidthForm{subitem}{\alph{mcitesubitemcount})}
\mciteSetBstSublistLabelBeginEnd{\mcitemaxwidthsubitemform\space}
{\relax}{\relax}

\bibitem{Gross:1973id}
D.~J. Gross and F.~Wilczek, \ifthenelse{\boolean{articletitles}}{\emph{{Ultraviolet behavior of nonabelian gauge theories}}, }{}\href{https://doi.org/10.1103/PhysRevLett.30.1343}{Phys.\ Rev.\ Lett.\  \textbf{30} (1973) 1343}\relax
\mciteBstWouldAddEndPuncttrue
\mciteSetBstMidEndSepPunct{\mcitedefaultmidpunct}
{\mcitedefaultendpunct}{\mcitedefaultseppunct}\relax
\EndOfBibitem
\bibitem{Politzer:1973fx}
H.~D. Politzer, \ifthenelse{\boolean{articletitles}}{\emph{{Reliable perturbative results for strong interactions?}}, }{}\href{https://doi.org/10.1103/PhysRevLett.30.1346}{Phys.\ Rev.\ Lett.\  \textbf{30} (1973) 1346}\relax
\mciteBstWouldAddEndPuncttrue
\mciteSetBstMidEndSepPunct{\mcitedefaultmidpunct}
{\mcitedefaultendpunct}{\mcitedefaultseppunct}\relax
\EndOfBibitem
\bibitem{Wilson:1974sk}
K.~G. Wilson, \ifthenelse{\boolean{articletitles}}{\emph{{Confinement of quarks}}, }{}\href{https://doi.org/10.1103/PhysRevD.10.2445}{Phys.\ Rev.\  \textbf{D10} (1974) 2445}\relax
\mciteBstWouldAddEndPuncttrue
\mciteSetBstMidEndSepPunct{\mcitedefaultmidpunct}
{\mcitedefaultendpunct}{\mcitedefaultseppunct}\relax
\EndOfBibitem
\bibitem{Nambu:1961tp}
Y.~Nambu and G.~Jona-Lasinio, \ifthenelse{\boolean{articletitles}}{\emph{{Dynamical model of elementary particles based on an analogy with superconductivity. \MakeUppercase{\romannumeral 1}}}, }{}\href{https://doi.org/10.1103/PhysRev.122.345}{Phys.\ Rev.\  \textbf{122} (1961) 345}\relax
\mciteBstWouldAddEndPuncttrue
\mciteSetBstMidEndSepPunct{\mcitedefaultmidpunct}
{\mcitedefaultendpunct}{\mcitedefaultseppunct}\relax
\EndOfBibitem
\bibitem{Gell-Mann:1968hlm}
M.~Gell-Mann, R.~J. Oakes, and B.~Renner, \ifthenelse{\boolean{articletitles}}{\emph{{Behavior of current divergences under $SU_{3} \times SU_{3}$}}, }{}\href{https://doi.org/10.1103/PhysRev.175.2195}{Phys.\ Rev.\  \textbf{175} (1968) 2195}\relax
\mciteBstWouldAddEndPuncttrue
\mciteSetBstMidEndSepPunct{\mcitedefaultmidpunct}
{\mcitedefaultendpunct}{\mcitedefaultseppunct}\relax
\EndOfBibitem
\bibitem{Roberts:2021nhw}
C.~D. Roberts, D.~G. Richards, T.~Horn, and L.~Chang, \ifthenelse{\boolean{articletitles}}{\emph{{Insights into the emergence of mass from studies of pion and kaon structure}}, }{}\href{https://doi.org/10.1016/j.ppnp.2021.103883}{Prog.\ Part.\ Nucl.\ Phys.\  \textbf{120} (2021) 103883}, \href{http://arxiv.org/abs/2102.01765}{{\normalfont\ttfamily arXiv:2102.01765}}\relax
\mciteBstWouldAddEndPuncttrue
\mciteSetBstMidEndSepPunct{\mcitedefaultmidpunct}
{\mcitedefaultendpunct}{\mcitedefaultseppunct}\relax
\EndOfBibitem
\bibitem{PDG2026}
Particle Data Group, F.~Takahashi {\em et~al.}, \ifthenelse{\boolean{articletitles}}{\emph{{\href{http://pdg.lbl.gov/}{Review of particle physics}}}, }{}\href{https://doi.org/10.1142/S0217751X26300115}{Mod.\ Phys.\  \textbf{A41} (2026) 2630011}\relax
\mciteBstWouldAddEndPuncttrue
\mciteSetBstMidEndSepPunct{\mcitedefaultmidpunct}
{\mcitedefaultendpunct}{\mcitedefaultseppunct}\relax
\EndOfBibitem
\bibitem{Brambilla:2010cs}
N.~Brambilla {\em et~al.}, \ifthenelse{\boolean{articletitles}}{\emph{{Heavy quarkonium: Progress, puzzles, and opportunities}}, }{}\href{https://doi.org/10.1140/epjc/s10052-010-1534-9}{Eur.\ Phys.\ J.\  \textbf{C71} (2011) 1534}, \href{http://arxiv.org/abs/1010.5827}{{\normalfont\ttfamily arXiv:1010.5827}}\relax
\mciteBstWouldAddEndPuncttrue
\mciteSetBstMidEndSepPunct{\mcitedefaultmidpunct}
{\mcitedefaultendpunct}{\mcitedefaultseppunct}\relax
\EndOfBibitem
\bibitem{Chen:2016qju}
H.-X. Chen, W.~Chen, X.~Liu, and S.-L. Zhu, \ifthenelse{\boolean{articletitles}}{\emph{{The hidden-charm pentaquark and tetraquark states}}, }{}\href{https://doi.org/10.1016/j.physrep.2016.05.004}{Phys.\ Rept.\  \textbf{639} (2016) 1}, \href{http://arxiv.org/abs/1601.02092}{{\normalfont\ttfamily arXiv:1601.02092}}\relax
\mciteBstWouldAddEndPuncttrue
\mciteSetBstMidEndSepPunct{\mcitedefaultmidpunct}
{\mcitedefaultendpunct}{\mcitedefaultseppunct}\relax
\EndOfBibitem
\bibitem{GellMann:1964nj}
M.~Gell-Mann, \ifthenelse{\boolean{articletitles}}{\emph{{A schematic model of baryons and mesons}}, }{}\href{https://doi.org/10.1016/S0031-9163(64)92001-3}{Phys.\ Lett.\  \textbf{8} (1964) 214}\relax
\mciteBstWouldAddEndPuncttrue
\mciteSetBstMidEndSepPunct{\mcitedefaultmidpunct}
{\mcitedefaultendpunct}{\mcitedefaultseppunct}\relax
\EndOfBibitem
\bibitem{Zweig:352337}
G.~Zweig,  \ifthenelse{\boolean{articletitles}}{\emph{{An SU$_3$ model for strong interaction symmetry and its breaking; Version~1}}}{}, \href{http://cds.cern.ch/record/352337}{CERN-TH-401}, CERN, Geneva, 1964\relax
\mciteBstWouldAddEndPuncttrue
\mciteSetBstMidEndSepPunct{\mcitedefaultmidpunct}
{\mcitedefaultendpunct}{\mcitedefaultseppunct}\relax
\EndOfBibitem
\bibitem{Petermann:1965qlk}
A.~Petermann, \ifthenelse{\boolean{articletitles}}{\emph{{Propriétés de l'étrangeté et une formule de masse pour les mésons vectoriels}}, }{}\href{https://doi.org/10.1016/0029-5582(65)90348-2}{Nucl.\ Phys.\  \textbf{63} (1965) 349}\relax
\mciteBstWouldAddEndPuncttrue
\mciteSetBstMidEndSepPunct{\mcitedefaultmidpunct}
{\mcitedefaultendpunct}{\mcitedefaultseppunct}\relax
\EndOfBibitem
\bibitem{Godfrey:1985xj}
S.~Godfrey and N.~Isgur, \ifthenelse{\boolean{articletitles}}{\emph{{Mesons in a relativized quark model with chromodynamics}}, }{}\href{https://doi.org/10.1103/PhysRevD.32.189}{Phys.\ Rev.\  \textbf{D32} (1985) 189}\relax
\mciteBstWouldAddEndPuncttrue
\mciteSetBstMidEndSepPunct{\mcitedefaultmidpunct}
{\mcitedefaultendpunct}{\mcitedefaultseppunct}\relax
\EndOfBibitem
\bibitem{Capstick:1986ter}
S.~Capstick and N.~Isgur, \ifthenelse{\boolean{articletitles}}{\emph{{Baryons in a relativized quark model with chromodynamics}}, }{}\href{https://doi.org/10.1103/physrevd.34.2809}{Phys.\ Rev.\  \textbf{D34} (1986) 2809}\relax
\mciteBstWouldAddEndPuncttrue
\mciteSetBstMidEndSepPunct{\mcitedefaultmidpunct}
{\mcitedefaultendpunct}{\mcitedefaultseppunct}\relax
\EndOfBibitem
\bibitem{BaBar:2003oey}
BaBar collaboration, B.~Aubert {\em et~al.}, \ifthenelse{\boolean{articletitles}}{\emph{{Observation of a narrow meson state decaying to $\Ds \piz$ at a mass of $2.32$\gevcc}}, }{}\href{https://doi.org/10.1103/PhysRevLett.90.242001}{Phys.\ Rev.\ Lett.\  \textbf{90} (2003) 242001}, \href{http://arxiv.org/abs/hep-ex/0304021}{{\normalfont\ttfamily arXiv:hep-ex/0304021}}\relax
\mciteBstWouldAddEndPuncttrue
\mciteSetBstMidEndSepPunct{\mcitedefaultmidpunct}
{\mcitedefaultendpunct}{\mcitedefaultseppunct}\relax
\EndOfBibitem
\bibitem{CLEO:2003ggt}
CLEO collaboration, D.~Besson {\em et~al.}, \ifthenelse{\boolean{articletitles}}{\emph{{Observation of a narrow resonance of mass $2.46\gevcc$ decaying to $D_\squark^{*+} \piz$ and confirmation of the $D^*_{sJ}(2317)$ state}}, }{}\href{https://doi.org/10.1103/PhysRevD.68.032002}{Phys.\ Rev.\  \textbf{D68} (2003) 032002}, Erratum \href{https://doi.org/10.1103/PhysRevD.75.119908}{ibid.\   \textbf{D75} (2007) 119908}, \href{http://arxiv.org/abs/hep-ex/0305100}{{\normalfont\ttfamily arXiv:hep-ex/0305100}}\relax
\mciteBstWouldAddEndPuncttrue
\mciteSetBstMidEndSepPunct{\mcitedefaultmidpunct}
{\mcitedefaultendpunct}{\mcitedefaultseppunct}\relax
\EndOfBibitem
\bibitem{Belle:2004dpt}
Belle collaboration, A.~Drutskoy {\em et~al.}, \ifthenelse{\boolean{articletitles}}{\emph{{Observation of $\decay{\Bdb}{ \D^*_{sJ}(2317)^+ \Km}$ decay}}, }{}\href{https://doi.org/10.1103/PhysRevLett.94.061802}{Phys.\ Rev.\ Lett.\  \textbf{94} (2005) 061802}, \href{http://arxiv.org/abs/hep-ex/0409026}{{\normalfont\ttfamily arXiv:hep-ex/0409026}}\relax
\mciteBstWouldAddEndPuncttrue
\mciteSetBstMidEndSepPunct{\mcitedefaultmidpunct}
{\mcitedefaultendpunct}{\mcitedefaultseppunct}\relax
\EndOfBibitem
\bibitem{Belle:2003guh}
Belle collaboration, P.~Krokovny {\em et~al.}, \ifthenelse{\boolean{articletitles}}{\emph{{Observation of the $D_{sJ}(2317)$ and $D_{sJ}(2457)$ in B decays}}, }{}\href{https://doi.org/10.1103/PhysRevLett.91.262002}{Phys.\ Rev.\ Lett.\  \textbf{91} (2003) 262002}, \href{http://arxiv.org/abs/hep-ex/0308019}{{\normalfont\ttfamily arXiv:hep-ex/0308019}}\relax
\mciteBstWouldAddEndPuncttrue
\mciteSetBstMidEndSepPunct{\mcitedefaultmidpunct}
{\mcitedefaultendpunct}{\mcitedefaultseppunct}\relax
\EndOfBibitem
\bibitem{BaBar:2004yux}
BaBar collaboration, B.~Aubert {\em et~al.}, \ifthenelse{\boolean{articletitles}}{\emph{{Study of $B \to D_{sJ}^{(*)+} \Dbar {\kern -0.35 cm {\phantom{D}}_{\phantom {sJ}}^{(*)}}$ decays}}, }{}\href{https://doi.org/10.1103/PhysRevLett.93.181801}{Phys.\ Rev.\ Lett.\  \textbf{93} (2004) 181801}, \href{http://arxiv.org/abs/hep-ex/0408041}{{\normalfont\ttfamily arXiv:hep-ex/0408041}}\relax
\mciteBstWouldAddEndPuncttrue
\mciteSetBstMidEndSepPunct{\mcitedefaultmidpunct}
{\mcitedefaultendpunct}{\mcitedefaultseppunct}\relax
\EndOfBibitem
\bibitem{Godfrey:2015dva}
S.~Godfrey and K.~Moats, \ifthenelse{\boolean{articletitles}}{\emph{{Properties of excited charm and charm-strange mesons}}, }{}\href{https://doi.org/10.1103/PhysRevD.93.034035}{Phys.\ Rev.\  \textbf{D93} (2016) 034035}, \href{http://arxiv.org/abs/1510.08305}{{\normalfont\ttfamily arXiv:1510.08305}}\relax
\mciteBstWouldAddEndPuncttrue
\mciteSetBstMidEndSepPunct{\mcitedefaultmidpunct}
{\mcitedefaultendpunct}{\mcitedefaultseppunct}\relax
\EndOfBibitem
\bibitem{Lahde:1999ih}
T.~A. L{\"a}hde, C.~J. Nyf{\"a}lt, and D.~O. Riska, \ifthenelse{\boolean{articletitles}}{\emph{{Spectra and M1 decay widths of heavy light mesons}}, }{}\href{https://doi.org/10.1016/S0375-9474(00)00154-8}{Nucl.\ Phys.\  \textbf{A674} (2000) 141}, \href{http://arxiv.org/abs/hep-ph/9908485}{{\normalfont\ttfamily arXiv:hep-ph/9908485}}\relax
\mciteBstWouldAddEndPuncttrue
\mciteSetBstMidEndSepPunct{\mcitedefaultmidpunct}
{\mcitedefaultendpunct}{\mcitedefaultseppunct}\relax
\EndOfBibitem
\bibitem{DiPierro:2001dwf}
M.~Di~Pierro and E.~Eichten, \ifthenelse{\boolean{articletitles}}{\emph{{Excited heavy-light systems and hadronic transitions}}, }{}\href{https://doi.org/10.1103/PhysRevD.64.114004}{Phys.\ Rev.\  \textbf{D64} (2001) 114004}, \href{http://arxiv.org/abs/hep-ph/0104208}{{\normalfont\ttfamily arXiv:hep-ph/0104208}}\relax
\mciteBstWouldAddEndPuncttrue
\mciteSetBstMidEndSepPunct{\mcitedefaultmidpunct}
{\mcitedefaultendpunct}{\mcitedefaultseppunct}\relax
\EndOfBibitem
\bibitem{Matsuki:2006zoi}
T.~Matsuki, T.~Morii, and K.~Sudoh, \ifthenelse{\boolean{articletitles}}{\emph{{New heavy-light mesons $Q \bar q$}}, }{}\href{https://doi.org/10.1143/PTP.117.1077}{Prog.\ Theor.\ Phys.\  \textbf{117} (2007) 1077}, \href{http://arxiv.org/abs/hep-ph/0605019}{{\normalfont\ttfamily arXiv:hep-ph/0605019}}\relax
\mciteBstWouldAddEndPuncttrue
\mciteSetBstMidEndSepPunct{\mcitedefaultmidpunct}
{\mcitedefaultendpunct}{\mcitedefaultseppunct}\relax
\EndOfBibitem
\bibitem{Lakhina:2006fy}
O.~Lakhina and E.~S. Swanson, \ifthenelse{\boolean{articletitles}}{\emph{{A canonical $\PD_s(2317)$?}}, }{}\href{https://doi.org/10.1016/j.physletb.2007.01.075}{Phys.\ Lett.\  \textbf{B650} (2007) 159}, \href{http://arxiv.org/abs/hep-ph/0608011}{{\normalfont\ttfamily arXiv:hep-ph/0608011}}\relax
\mciteBstWouldAddEndPuncttrue
\mciteSetBstMidEndSepPunct{\mcitedefaultmidpunct}
{\mcitedefaultendpunct}{\mcitedefaultseppunct}\relax
\EndOfBibitem
\bibitem{WooLee:2006kdh}
I.~Woo~Lee and T.~Lee, \ifthenelse{\boolean{articletitles}}{\emph{{Why there is no spin-orbit inversion in heavy-light mesons?}}, }{}\href{https://doi.org/10.1103/PhysRevD.76.014017}{Phys.\ Rev.\  \textbf{D76} (2007) 014017}, \href{http://arxiv.org/abs/hep-ph/0612345}{{\normalfont\ttfamily arXiv:hep-ph/0612345}}\relax
\mciteBstWouldAddEndPuncttrue
\mciteSetBstMidEndSepPunct{\mcitedefaultmidpunct}
{\mcitedefaultendpunct}{\mcitedefaultseppunct}\relax
\EndOfBibitem
\bibitem{Koponen:2007nr}
UKQCD collaboration, J.~Koponen, \ifthenelse{\boolean{articletitles}}{\emph{{Energies of $\PB_\squark$ meson excited states: A lattice study}}, }{}\href{https://doi.org/10.1103/PhysRevD.78.074509}{Phys.\ Rev.\  \textbf{D78} (2008) 074509}, \href{http://arxiv.org/abs/0708.2807}{{\normalfont\ttfamily arXiv:0708.2807}}\relax
\mciteBstWouldAddEndPuncttrue
\mciteSetBstMidEndSepPunct{\mcitedefaultmidpunct}
{\mcitedefaultendpunct}{\mcitedefaultseppunct}\relax
\EndOfBibitem
\bibitem{Gregory:2010gm}
E.~B. Gregory {\em et~al.}, \ifthenelse{\boolean{articletitles}}{\emph{{Precise $\PB, \PB_\squark$ and $\PB_\cquark$ meson spectroscopy from full lattice QCD}}, }{}\href{https://doi.org/10.1103/PhysRevD.83.014506}{Phys.\ Rev.\  \textbf{D83} (2011) 014506}, \href{http://arxiv.org/abs/1010.3848}{{\normalfont\ttfamily arXiv:1010.3848}}\relax
\mciteBstWouldAddEndPuncttrue
\mciteSetBstMidEndSepPunct{\mcitedefaultmidpunct}
{\mcitedefaultendpunct}{\mcitedefaultseppunct}\relax
\EndOfBibitem
\bibitem{Lang:2015hza}
C.~B. Lang, D.~Mohler, S.~Prelovsek, and R.~M. Woloshyn, \ifthenelse{\boolean{articletitles}}{\emph{{Predicting positive parity $\PB_\squark$ mesons from lattice QCD}}, }{}\href{https://doi.org/10.1016/j.physletb.2015.08.038}{Phys.\ Lett.\  \textbf{B750} (2015) 17}, \href{http://arxiv.org/abs/1501.01646}{{\normalfont\ttfamily arXiv:1501.01646}}\relax
\mciteBstWouldAddEndPuncttrue
\mciteSetBstMidEndSepPunct{\mcitedefaultmidpunct}
{\mcitedefaultendpunct}{\mcitedefaultseppunct}\relax
\EndOfBibitem
\bibitem{Wurtz:2015mqa}
M.~Wurtz, R.~Lewis, and R.~M. Woloshyn, \ifthenelse{\boolean{articletitles}}{\emph{{Free-form smearing for bottomonium and \PB meson spectroscopy}}, }{}\href{https://doi.org/10.1103/PhysRevD.92.054504}{Phys.\ Rev.\  \textbf{D92} (2015) 054504}, \href{http://arxiv.org/abs/1505.04410}{{\normalfont\ttfamily arXiv:1505.04410}}\relax
\mciteBstWouldAddEndPuncttrue
\mciteSetBstMidEndSepPunct{\mcitedefaultmidpunct}
{\mcitedefaultendpunct}{\mcitedefaultseppunct}\relax
\EndOfBibitem
\bibitem{Hudspith:2023loy}
R.~J. Hudspith and D.~Mohler, \ifthenelse{\boolean{articletitles}}{\emph{{Exotic tetraquark states with two \bquarkbar quarks and $J^P=0^+$ and $1^+$ $\PB_\squark$ states in a nonperturbatively tuned lattice NRQCD setup}}, }{}\href{https://doi.org/10.1103/PhysRevD.107.114510}{Phys.\ Rev.\  \textbf{D107} (2023) 114510}, \href{http://arxiv.org/abs/2303.17295}{{\normalfont\ttfamily arXiv:2303.17295}}\relax
\mciteBstWouldAddEndPuncttrue
\mciteSetBstMidEndSepPunct{\mcitedefaultmidpunct}
{\mcitedefaultendpunct}{\mcitedefaultseppunct}\relax
\EndOfBibitem
\bibitem{Gayer:2024akw}
Hadron Spectrum collaboration, L.~Gayer, S.~M. Ryan, and D.~J. Wilson, \ifthenelse{\boolean{articletitles}}{\emph{{Highly excited \PB, $\PB_{\squark}$ and $\PB_{\cquark}$ meson spectroscopy from lattice QCD}}, }{}\href{https://doi.org/10.1007/JHEP01(2025)123}{JHEP \textbf{01} (2025) 123}, \href{http://arxiv.org/abs/2408.02126}{{\normalfont\ttfamily arXiv:2408.02126}}\relax
\mciteBstWouldAddEndPuncttrue
\mciteSetBstMidEndSepPunct{\mcitedefaultmidpunct}
{\mcitedefaultendpunct}{\mcitedefaultseppunct}\relax
\EndOfBibitem
\bibitem{Liu:2012zya}
L.~Liu {\em et~al.}, \ifthenelse{\boolean{articletitles}}{\emph{{Interactions of charmed mesons with light pseudoscalar mesons from lattice QCD and implications on the nature of the $D_{s0}^*(2317)$}}, }{}\href{https://doi.org/10.1103/PhysRevD.87.014508}{Phys.\ Rev.\  \textbf{D87} (2013) 014508}, \href{http://arxiv.org/abs/1208.4535}{{\normalfont\ttfamily arXiv:1208.4535}}\relax
\mciteBstWouldAddEndPuncttrue
\mciteSetBstMidEndSepPunct{\mcitedefaultmidpunct}
{\mcitedefaultendpunct}{\mcitedefaultseppunct}\relax
\EndOfBibitem
\bibitem{Bardeen:2003kt}
W.~A. Bardeen, E.~J. Eichten, and C.~T. Hill, \ifthenelse{\boolean{articletitles}}{\emph{{Chiral multiplets of heavy-light mesons}}, }{}\href{https://doi.org/10.1103/PhysRevD.68.054024}{Phys.\ Rev.\  \textbf{D68} (2003) 054024}, \href{http://arxiv.org/abs/hep-ph/0305049}{{\normalfont\ttfamily arXiv:hep-ph/0305049}}\relax
\mciteBstWouldAddEndPuncttrue
\mciteSetBstMidEndSepPunct{\mcitedefaultmidpunct}
{\mcitedefaultendpunct}{\mcitedefaultseppunct}\relax
\EndOfBibitem
\bibitem{Nowak:2003ra}
M.~A. Nowak, M.~Rho, and I.~Zahed, \ifthenelse{\boolean{articletitles}}{\emph{{Chiral doubling of heavy light hadrons: \babar $2317\mevcc$ and \cleo $2463\mevcc$ discoveries}}, }{}\href{https://doi.org/https://www.actaphys.uj.edu.pl/R/35/10/2377/pdf}{{\href{https://wwwXXXactaphysXXXujXXXeduXXXpl/R/35/10/2377/pdf}{Acta Phys.\ Polon.\ \!\!}} \textbf{\href{https://www.actaphys.uj.edu.pl/R/35/10/2377/pdf}{B35}} (\href{https://www.actaphys.uj.edu.pl/R/35/10/2377/pdf}{2004}) \href{https://www.actaphys.uj.edu.pl/R/35/10/2377/pdf}{2377}}, \href{http://arxiv.org/abs/hep-ph/0307102}{{\normalfont\ttfamily arXiv:hep-ph/0307102}}\relax
\mciteBstWouldAddEndPuncttrue
\mciteSetBstMidEndSepPunct{\mcitedefaultmidpunct}
{\mcitedefaultendpunct}{\mcitedefaultseppunct}\relax
\EndOfBibitem
\bibitem{Colangelo:2005gb}
P.~Colangelo, F.~De~Fazio, and R.~Ferrandes, \ifthenelse{\boolean{articletitles}}{\emph{{Bounding effective parameters in the chiral Lagrangian for excited heavy mesons}}, }{}\href{https://doi.org/10.1016/j.physletb.2006.01.021}{Phys.\ Lett.\  \textbf{B634} (2006) 235}, \href{http://arxiv.org/abs/hep-ph/0511317}{{\normalfont\ttfamily arXiv:hep-ph/0511317}}\relax
\mciteBstWouldAddEndPuncttrue
\mciteSetBstMidEndSepPunct{\mcitedefaultmidpunct}
{\mcitedefaultendpunct}{\mcitedefaultseppunct}\relax
\EndOfBibitem
\bibitem{Badalian:2007yr}
A.~M. Badalian, Y.~A. Simonov, and M.~A. Trusov, \ifthenelse{\boolean{articletitles}}{\emph{{Chiral transitions in heavy-light mesons}}, }{}\href{https://doi.org/10.1103/PhysRevD.77.074017}{Phys.\ Rev.\  \textbf{D77} (2008) 074017}, \href{http://arxiv.org/abs/0712.3943}{{\normalfont\ttfamily arXiv:0712.3943}}\relax
\mciteBstWouldAddEndPuncttrue
\mciteSetBstMidEndSepPunct{\mcitedefaultmidpunct}
{\mcitedefaultendpunct}{\mcitedefaultseppunct}\relax
\EndOfBibitem
\bibitem{Colangelo:2012xi}
P.~Colangelo, F.~De~Fazio, F.~Giannuzzi, and S.~Nicotri, \ifthenelse{\boolean{articletitles}}{\emph{{New meson spectroscopy with open charm and beauty}}, }{}\href{https://doi.org/10.1103/PhysRevD.86.054024}{Phys.\ Rev.\  \textbf{D86} (2012) 054024}, \href{http://arxiv.org/abs/1207.6940}{{\normalfont\ttfamily arXiv:1207.6940}}\relax
\mciteBstWouldAddEndPuncttrue
\mciteSetBstMidEndSepPunct{\mcitedefaultmidpunct}
{\mcitedefaultendpunct}{\mcitedefaultseppunct}\relax
\EndOfBibitem
\bibitem{Wang:2015mxa}
Z.-G. Wang, \ifthenelse{\boolean{articletitles}}{\emph{{Analysis of the masses and decay constants of the heavy-light mesons with QCD sum rules}}, }{}\href{https://doi.org/10.1140/epjc/s10052-015-3653-9}{Eur.\ Phys.\ J.\  \textbf{C75} (2015) 427}, \href{http://arxiv.org/abs/1506.01993}{{\normalfont\ttfamily arXiv:1506.01993}}\relax
\mciteBstWouldAddEndPuncttrue
\mciteSetBstMidEndSepPunct{\mcitedefaultmidpunct}
{\mcitedefaultendpunct}{\mcitedefaultseppunct}\relax
\EndOfBibitem
\bibitem{Alhakami:2020vil}
M.~H. Alhakami, \ifthenelse{\boolean{articletitles}}{\emph{{Predictions for the beauty meson spectrum}}, }{}\href{https://doi.org/10.1103/PhysRevD.103.034009}{Phys.\ Rev.\  \textbf{D103} (2021) 034009}, \href{http://arxiv.org/abs/2006.16878}{{\normalfont\ttfamily arXiv:2006.16878}}\relax
\mciteBstWouldAddEndPuncttrue
\mciteSetBstMidEndSepPunct{\mcitedefaultmidpunct}
{\mcitedefaultendpunct}{\mcitedefaultseppunct}\relax
\EndOfBibitem
\bibitem{Gandhi:2022nnk}
K.~Gandhi and A.~K. Rai, \ifthenelse{\boolean{articletitles}}{\emph{{Study of $\B$, $\B_\squark$ mesons using heavy quark effective theory}}, }{}\href{https://doi.org/10.1140/epjc/s10052-022-10719-w}{Eur.\ Phys.\ J.\  \textbf{C82} (2022) 777}, \href{http://arxiv.org/abs/2208.08791}{{\normalfont\ttfamily arXiv:2208.08791}}\relax
\mciteBstWouldAddEndPuncttrue
\mciteSetBstMidEndSepPunct{\mcitedefaultmidpunct}
{\mcitedefaultendpunct}{\mcitedefaultseppunct}\relax
\EndOfBibitem
\bibitem{Wang:2007tu}
Z.-G. Wang, \ifthenelse{\boolean{articletitles}}{\emph{{Reanalysis of the $(0^+, 1^+)$ states $B_{s0}$ and $B_{s1}$ with QCD sum rules}}, }{}\href{https://doi.org/10.1088/0256-307X/25/11/020}{Chin.\ Phys.\ Lett.\  \textbf{25} (2008) 3908}, \href{http://arxiv.org/abs/0712.0118}{{\normalfont\ttfamily arXiv:0712.0118}}\relax
\mciteBstWouldAddEndPuncttrue
\mciteSetBstMidEndSepPunct{\mcitedefaultmidpunct}
{\mcitedefaultendpunct}{\mcitedefaultseppunct}\relax
\EndOfBibitem
\bibitem{Vishwakarma:2022sly}
K.~K. Vishwakarma and A.~Upadhyay, \ifthenelse{\boolean{articletitles}}{\emph{{Bottom mesons $J^P$ = $0^-$, $1^-$ and $0^+$, $1^+$ in heavy hadron chiral perturbation theory}}, }{}\href{https://doi.org/10.1140/epja/s10050-022-00680-3}{Eur.\ Phys.\ J.\  \textbf{A58} (2022) 38}\relax
\mciteBstWouldAddEndPuncttrue
\mciteSetBstMidEndSepPunct{\mcitedefaultmidpunct}
{\mcitedefaultendpunct}{\mcitedefaultseppunct}\relax
\EndOfBibitem
\bibitem{Kolomeitsev:2003ac}
E.~E. Kolomeitsev and M.~F.~M. Lutz, \ifthenelse{\boolean{articletitles}}{\emph{{On heavy-light meson resonances and chiral symmetry}}, }{}\href{https://doi.org/10.1016/j.physletb.2003.10.118}{Phys.\ Lett.\  \textbf{B582} (2004) 39}, \href{http://arxiv.org/abs/hep-ph/0307133}{{\normalfont\ttfamily arXiv:hep-ph/0307133}}\relax
\mciteBstWouldAddEndPuncttrue
\mciteSetBstMidEndSepPunct{\mcitedefaultmidpunct}
{\mcitedefaultendpunct}{\mcitedefaultseppunct}\relax
\EndOfBibitem
\bibitem{Barnes:2003dj}
T.~Barnes, F.~E. Close, and H.~J. Lipkin, \ifthenelse{\boolean{articletitles}}{\emph{{Implications of a DK molecule at $2.32$\gev}}, }{}\href{https://doi.org/10.1103/PhysRevD.68.054006}{Phys.\ Rev.\  \textbf{D68} (2003) 054006}, \href{http://arxiv.org/abs/hep-ph/0305025}{{\normalfont\ttfamily arXiv:hep-ph/0305025}}\relax
\mciteBstWouldAddEndPuncttrue
\mciteSetBstMidEndSepPunct{\mcitedefaultmidpunct}
{\mcitedefaultendpunct}{\mcitedefaultseppunct}\relax
\EndOfBibitem
\bibitem{Guo:2006fu}
F.-K. Guo {\em et~al.}, \ifthenelse{\boolean{articletitles}}{\emph{{Dynamically generated $0^+$ heavy mesons in a heavy chiral unitary approach}}, }{}\href{https://doi.org/10.1016/j.physletb.2006.08.064}{Phys.\ Lett.\  \textbf{B641} (2006) 278}, \href{http://arxiv.org/abs/hep-ph/0603072}{{\normalfont\ttfamily arXiv:hep-ph/0603072}}\relax
\mciteBstWouldAddEndPuncttrue
\mciteSetBstMidEndSepPunct{\mcitedefaultmidpunct}
{\mcitedefaultendpunct}{\mcitedefaultseppunct}\relax
\EndOfBibitem
\bibitem{Altenbuchinger:2013vwa}
M.~Altenbuchinger, L.-S. Geng, and W.~Weise, \ifthenelse{\boolean{articletitles}}{\emph{{Scattering lengths of Nambu--Goldstone bosons off \PD mesons and dynamically generated heavy-light mesons}}, }{}\href{https://doi.org/10.1103/PhysRevD.89.014026}{Phys.\ Rev.\  \textbf{D89} (2014) 014026}, \href{http://arxiv.org/abs/1309.4743}{{\normalfont\ttfamily arXiv:1309.4743}}\relax
\mciteBstWouldAddEndPuncttrue
\mciteSetBstMidEndSepPunct{\mcitedefaultmidpunct}
{\mcitedefaultendpunct}{\mcitedefaultseppunct}\relax
\EndOfBibitem
\bibitem{Ortega:2016pgg}
P.~G. Ortega, J.~Segovia, D.~R. Entem, and F.~Fern{\'a}ndez, \ifthenelse{\boolean{articletitles}}{\emph{{Threshold effects in P-wave bottom-strange mesons}}, }{}\href{https://doi.org/10.1103/PhysRevD.95.034010}{Phys.\ Rev.\  \textbf{D95} (2017) 034010}, \href{http://arxiv.org/abs/1612.04826}{{\normalfont\ttfamily arXiv:1612.04826}}\relax
\mciteBstWouldAddEndPuncttrue
\mciteSetBstMidEndSepPunct{\mcitedefaultmidpunct}
{\mcitedefaultendpunct}{\mcitedefaultseppunct}\relax
\EndOfBibitem
\bibitem{Albaladejo:2016ztm}
M.~Albaladejo, P.~Fernandez-Soler, J.~Nieves, and P.~G. Ortega, \ifthenelse{\boolean{articletitles}}{\emph{{Lowest-lying even-parity ${\overline{B}}_\squark$ mesons: Heavy-quark spin-flavor symmetry, chiral dynamics, and constituent quark-model bare masses}}, }{}\href{https://doi.org/10.1140/epjc/s10052-017-4735-7}{Eur.\ Phys.\ J.\  \textbf{C77} (2017) 170}, \href{http://arxiv.org/abs/1612.07782}{{\normalfont\ttfamily arXiv:1612.07782}}\relax
\mciteBstWouldAddEndPuncttrue
\mciteSetBstMidEndSepPunct{\mcitedefaultmidpunct}
{\mcitedefaultendpunct}{\mcitedefaultseppunct}\relax
\EndOfBibitem
\bibitem{Zhou:2020moj}
Z.-Y. Zhou and Z.~Xiao, \ifthenelse{\boolean{articletitles}}{\emph{{Two-pole structures in a relativistic Friedrichs--Lee-QPC scheme}}, }{}\href{https://doi.org/10.1140/epjc/s10052-021-09329-9}{Eur.\ Phys.\ J.\  \textbf{C81} (2021) 551}, \href{http://arxiv.org/abs/2008.08002}{{\normalfont\ttfamily arXiv:2008.08002}}\relax
\mciteBstWouldAddEndPuncttrue
\mciteSetBstMidEndSepPunct{\mcitedefaultmidpunct}
{\mcitedefaultendpunct}{\mcitedefaultseppunct}\relax
\EndOfBibitem
\bibitem{Guo:2021rjv}
X.-Y. Guo and M.~F.~M. Lutz, \ifthenelse{\boolean{articletitles}}{\emph{{Chiral excitations of open-beauty systems}}, }{}\href{https://doi.org/10.1103/PhysRevD.104.054035}{Phys.\ Rev.\  \textbf{D104} (2021) 054035}, \href{http://arxiv.org/abs/2103.11323}{{\normalfont\ttfamily arXiv:2103.11323}}\relax
\mciteBstWouldAddEndPuncttrue
\mciteSetBstMidEndSepPunct{\mcitedefaultmidpunct}
{\mcitedefaultendpunct}{\mcitedefaultseppunct}\relax
\EndOfBibitem
\bibitem{Ni:2023lvx}
R.-H. Ni, J.-J. Wu, and X.-H. Zhong, \ifthenelse{\boolean{articletitles}}{\emph{{Unified unquenched quark model for heavy-light mesons with chiral dynamics}}, }{}\href{https://doi.org/10.1103/PhysRevD.109.116006}{Phys.\ Rev.\  \textbf{D109} (2024) 116006}, \href{http://arxiv.org/abs/2312.04765}{{\normalfont\ttfamily arXiv:2312.04765}}\relax
\mciteBstWouldAddEndPuncttrue
\mciteSetBstMidEndSepPunct{\mcitedefaultmidpunct}
{\mcitedefaultendpunct}{\mcitedefaultseppunct}\relax
\EndOfBibitem
\bibitem{Zhang:2024usz}
Z.-L. Zhang {\em et~al.}, \ifthenelse{\boolean{articletitles}}{\emph{{Masses and radiative decay widths of $D_{s0}^*(2317)$ and $D_{s1}^\prime(2460)$ and their bottom analogs}}, }{}\href{https://doi.org/10.1103/PhysRevD.110.094037}{Phys.\ Rev.\  \textbf{D110} (2024) 094037}, \href{http://arxiv.org/abs/2409.05337}{{\normalfont\ttfamily arXiv:2409.05337}}\relax
\mciteBstWouldAddEndPuncttrue
\mciteSetBstMidEndSepPunct{\mcitedefaultmidpunct}
{\mcitedefaultendpunct}{\mcitedefaultseppunct}\relax
\EndOfBibitem
\bibitem{Cleven:2010aw}
M.~Cleven, F.-K. Guo, C.~Hanhart, and U.-G. Mei{\ss}ner, \ifthenelse{\boolean{articletitles}}{\emph{{Light meson mass dependence of the positive-parity heavy-strange mesons}}, }{}\href{https://doi.org/10.1140/epja/i2011-11019-2}{Eur.\ Phys.\ J.\  \textbf{A47} (2011) 19}, \href{http://arxiv.org/abs/1009.3804}{{\normalfont\ttfamily arXiv:1009.3804}}\relax
\mciteBstWouldAddEndPuncttrue
\mciteSetBstMidEndSepPunct{\mcitedefaultmidpunct}
{\mcitedefaultendpunct}{\mcitedefaultseppunct}\relax
\EndOfBibitem
\bibitem{Fu:2021wde}
H.-L. Fu {\em et~al.}, \ifthenelse{\boolean{articletitles}}{\emph{{Update on strong and radiative decays of the $D_{s0}^*(2317)$ and $D_{s1}(2460)$ and their bottom cousins}}, }{}\href{https://doi.org/10.1140/epja/s10050-022-00724-8}{Eur.\ Phys.\ J.\  \textbf{A58} (2022) 70}, \href{http://arxiv.org/abs/2111.09481}{{\normalfont\ttfamily arXiv:2111.09481}}\relax
\mciteBstWouldAddEndPuncttrue
\mciteSetBstMidEndSepPunct{\mcitedefaultmidpunct}
{\mcitedefaultendpunct}{\mcitedefaultseppunct}\relax
\EndOfBibitem
\bibitem{Vijande:2007ke}
J.~Vijande, A.~Valcarce, and F.~Fern{\'a}ndez, \ifthenelse{\boolean{articletitles}}{\emph{{$\PB$ meson spectroscopy}}, }{}\href{https://doi.org/10.1103/PhysRevD.77.017501}{Phys.\ Rev.\  \textbf{D77} (2008) 017501}, \href{http://arxiv.org/abs/0711.2359}{{\normalfont\ttfamily arXiv:0711.2359}}\relax
\mciteBstWouldAddEndPuncttrue
\mciteSetBstMidEndSepPunct{\mcitedefaultmidpunct}
{\mcitedefaultendpunct}{\mcitedefaultseppunct}\relax
\EndOfBibitem
\bibitem{Torres-Rincon:2014ffa}
J.~M. Torres-Rincon, L.~Tolos, and O.~Romanets, \ifthenelse{\boolean{articletitles}}{\emph{{Open bottom states and the $\bar B$-meson propagation in hadronic matter}}, }{}\href{https://doi.org/10.1103/PhysRevD.89.074042}{Phys.\ Rev.\  \textbf{D89} (2014) 074042}, \href{http://arxiv.org/abs/1403.1371}{{\normalfont\ttfamily arXiv:1403.1371}}\relax
\mciteBstWouldAddEndPuncttrue
\mciteSetBstMidEndSepPunct{\mcitedefaultmidpunct}
{\mcitedefaultendpunct}{\mcitedefaultseppunct}\relax
\EndOfBibitem
\bibitem{Du:2017zvv}
M.-L. Du {\em et~al.}, \ifthenelse{\boolean{articletitles}}{\emph{{Towards a new paradigm for heavy-light meson spectroscopy}}, }{}\href{https://doi.org/10.1103/PhysRevD.98.094018}{Phys.\ Rev.\  \textbf{D98} (2018) 094018}, \href{http://arxiv.org/abs/1712.07957}{{\normalfont\ttfamily arXiv:1712.07957}}\relax
\mciteBstWouldAddEndPuncttrue
\mciteSetBstMidEndSepPunct{\mcitedefaultmidpunct}
{\mcitedefaultendpunct}{\mcitedefaultseppunct}\relax
\EndOfBibitem
\bibitem{Yang:2022vdb}
Z.~Yang {\em et~al.}, \ifthenelse{\boolean{articletitles}}{\emph{{The investigations of the P-wave B$_{s}$ states combining quark model and lattice QCD in the coupled channel framework}}, }{}\href{https://doi.org/10.1007/JHEP01(2023)058}{JHEP \textbf{01} (2023) 058}, \href{http://arxiv.org/abs/2207.07320}{{\normalfont\ttfamily arXiv:2207.07320}}\relax
\mciteBstWouldAddEndPuncttrue
\mciteSetBstMidEndSepPunct{\mcitedefaultmidpunct}
{\mcitedefaultendpunct}{\mcitedefaultseppunct}\relax
\EndOfBibitem
\bibitem{Kim:2023htt}
H.-J. Kim and H.-C. Kim, \ifthenelse{\boolean{articletitles}}{\emph{{$D^*_{s0}(2317)$ and $B^*_{s0}$ as molecular states}}, }{}\href{https://doi.org/10.1093/ptep/ptae095}{PTEP \textbf{2024} (2024) 073D01}, \href{http://arxiv.org/abs/2310.13370}{{\normalfont\ttfamily arXiv:2310.13370}}\relax
\mciteBstWouldAddEndPuncttrue
\mciteSetBstMidEndSepPunct{\mcitedefaultmidpunct}
{\mcitedefaultendpunct}{\mcitedefaultseppunct}\relax
\EndOfBibitem
\bibitem{Cheng:2003kg}
H.-Y. Cheng and W.-S. Hou, \ifthenelse{\boolean{articletitles}}{\emph{{B decays as spectroscope for charmed four quark states}}, }{}\href{https://doi.org/10.1016/S0370-2693(03)00834-7}{Phys.\ Lett.\  \textbf{B566} (2003) 193}, \href{http://arxiv.org/abs/hep-ph/0305038}{{\normalfont\ttfamily arXiv:hep-ph/0305038}}\relax
\mciteBstWouldAddEndPuncttrue
\mciteSetBstMidEndSepPunct{\mcitedefaultmidpunct}
{\mcitedefaultendpunct}{\mcitedefaultseppunct}\relax
\EndOfBibitem
\bibitem{Maiani:2004vq}
L.~Maiani, F.~Piccinini, A.~D. Polosa, and V.~Riquer, \ifthenelse{\boolean{articletitles}}{\emph{{Diquark-antidiquarks with hidden or open charm and the nature of $X(3872)$}}, }{}\href{https://doi.org/10.1103/PhysRevD.71.014028}{Phys.\ Rev.\  \textbf{D71} (2005) 014028}, \href{http://arxiv.org/abs/hep-ph/0412098}{{\normalfont\ttfamily arXiv:hep-ph/0412098}}\relax
\mciteBstWouldAddEndPuncttrue
\mciteSetBstMidEndSepPunct{\mcitedefaultmidpunct}
{\mcitedefaultendpunct}{\mcitedefaultseppunct}\relax
\EndOfBibitem
\bibitem{Dmitrasinovic:2012zz}
V.~Dmitrasinovic, \ifthenelse{\boolean{articletitles}}{\emph{{Chiral symmetry of heavy-light scalar mesons with $U_A(1)$ symmetry breaking}}, }{}\href{https://doi.org/10.1103/PhysRevD.86.016006}{Phys.\ Rev.\  \textbf{D86} (2012) 016006}\relax
\mciteBstWouldAddEndPuncttrue
\mciteSetBstMidEndSepPunct{\mcitedefaultmidpunct}
{\mcitedefaultendpunct}{\mcitedefaultseppunct}\relax
\EndOfBibitem
\bibitem{Olsen:2017bmm}
S.~L. Olsen, T.~Skwarnicki, and D.~Zieminska, \ifthenelse{\boolean{articletitles}}{\emph{{Nonstandard heavy mesons and baryons: Experimental evidence}}, }{}\href{https://doi.org/10.1103/RevModPhys.90.015003}{Rev.\ Mod.\ Phys.\  \textbf{90} (2018) 015003}, \href{http://arxiv.org/abs/1708.04012}{{\normalfont\ttfamily arXiv:1708.04012}}\relax
\mciteBstWouldAddEndPuncttrue
\mciteSetBstMidEndSepPunct{\mcitedefaultmidpunct}
{\mcitedefaultendpunct}{\mcitedefaultseppunct}\relax
\EndOfBibitem
\bibitem{Guo:2017jvc}
F.-K. Guo {\em et~al.}, \ifthenelse{\boolean{articletitles}}{\emph{{Hadronic molecules}}, }{}\href{https://doi.org/10.1103/RevModPhys.90.015004}{Rev.\ Mod.\ Phys.\  \textbf{90} (2018) 015004}, Erratum \href{https://doi.org/10.1103/RevModPhys.94.029901}{ibid.\   \textbf{94} (2022) 029901}, \href{http://arxiv.org/abs/1705.00141}{{\normalfont\ttfamily arXiv:1705.00141}}\relax
\mciteBstWouldAddEndPuncttrue
\mciteSetBstMidEndSepPunct{\mcitedefaultmidpunct}
{\mcitedefaultendpunct}{\mcitedefaultseppunct}\relax
\EndOfBibitem
\bibitem{Karliner:2017qhf}
M.~Karliner, J.~L. Rosner, and T.~Skwarnicki, \ifthenelse{\boolean{articletitles}}{\emph{{Multiquark states}}, }{}\href{https://doi.org/10.1146/annurev-nucl-101917-020902}{Ann.\ Rev.\ Nucl.\ Part.\ Sci.\  \textbf{68} (2018) 17}, \href{http://arxiv.org/abs/1711.10626}{{\normalfont\ttfamily arXiv:1711.10626}}\relax
\mciteBstWouldAddEndPuncttrue
\mciteSetBstMidEndSepPunct{\mcitedefaultmidpunct}
{\mcitedefaultendpunct}{\mcitedefaultseppunct}\relax
\EndOfBibitem
\bibitem{Liu:2019zoy}
Y.-R. Liu {\em et~al.}, \ifthenelse{\boolean{articletitles}}{\emph{{Pentaquark and tetraquark states}}, }{}\href{https://doi.org/10.1016/j.ppnp.2019.04.003}{Prog.\ Part.\ Nucl.\ Phys.\  \textbf{107} (2019) 237}, \href{http://arxiv.org/abs/1903.11976}{{\normalfont\ttfamily arXiv:1903.11976}}\relax
\mciteBstWouldAddEndPuncttrue
\mciteSetBstMidEndSepPunct{\mcitedefaultmidpunct}
{\mcitedefaultendpunct}{\mcitedefaultseppunct}\relax
\EndOfBibitem
\bibitem{Wise:1992hn}
M.~B. Wise, \ifthenelse{\boolean{articletitles}}{\emph{{Chiral perturbation theory for hadrons containing a heavy quark}}, }{}\href{https://doi.org/10.1103/PhysRevD.45.R2188}{Phys.\ Rev.\  \textbf{D45} (1992) R2188}\relax
\mciteBstWouldAddEndPuncttrue
\mciteSetBstMidEndSepPunct{\mcitedefaultmidpunct}
{\mcitedefaultendpunct}{\mcitedefaultseppunct}\relax
\EndOfBibitem
\bibitem{Yan:1992gz}
T.-M. Yan {\em et~al.}, \ifthenelse{\boolean{articletitles}}{\emph{{Heavy quark symmetry and chiral dynamics}}, }{}\href{https://doi.org/10.1103/PhysRevD.46.1148}{Phys.\ Rev.\  \textbf{D46} (1992) 1148}, Erratum \href{https://doi.org/10.1103/PhysRevD.55.5851}{ibid.\   \textbf{D55} (1997) 5851}\relax
\mciteBstWouldAddEndPuncttrue
\mciteSetBstMidEndSepPunct{\mcitedefaultmidpunct}
{\mcitedefaultendpunct}{\mcitedefaultseppunct}\relax
\EndOfBibitem
\bibitem{Cho:1994zu}
P.~L. Cho and M.~B. Wise, \ifthenelse{\boolean{articletitles}}{\emph{{Remarks on \decay{ \PD_\squark^*}{\PD^{}_\squark \piz} decay}}, }{}\href{https://doi.org/10.1103/PhysRevD.49.6228}{Phys.\ Rev.\  \textbf{D49} (1994) 6228}, \href{http://arxiv.org/abs/hep-ph/9401301}{{\normalfont\ttfamily arXiv:hep-ph/9401301}}\relax
\mciteBstWouldAddEndPuncttrue
\mciteSetBstMidEndSepPunct{\mcitedefaultmidpunct}
{\mcitedefaultendpunct}{\mcitedefaultseppunct}\relax
\EndOfBibitem
\bibitem{D0:2016mwd}
D0 collaboration, V.~M. Abazov {\em et~al.}, \ifthenelse{\boolean{articletitles}}{\emph{{Evidence for a $B_s^0 \pi^\pm$ state}}, }{}\href{https://doi.org/10.1103/PhysRevLett.117.022003}{Phys.\ Rev.\ Lett.\  \textbf{117} (2016) 022003}, \href{http://arxiv.org/abs/1602.07588}{{\normalfont\ttfamily arXiv:1602.07588}}\relax
\mciteBstWouldAddEndPuncttrue
\mciteSetBstMidEndSepPunct{\mcitedefaultmidpunct}
{\mcitedefaultendpunct}{\mcitedefaultseppunct}\relax
\EndOfBibitem
\bibitem{LHCB-PAPER-2016-029}
LHCb collaboration, R.~Aaij {\em et~al.}, \ifthenelse{\boolean{articletitles}}{\emph{{Search for structure in the $\Bs\pipm$ invariant mass spectrum}}, }{}\href{https://doi.org/10.1103/PhysRevLett.117.152003}{Phys.\ ~Rev.\ ~Lett.\  \textbf{117} (2016) 152003}, Addendum \href{https://doi.org/10.1103/PhysRevLett.118.109904}{ibid.\   \textbf{118} (2017) 109904}, \href{http://arxiv.org/abs/1608.00435}{{\normalfont\ttfamily arXiv:1608.00435}}\relax
\mciteBstWouldAddEndPuncttrue
\mciteSetBstMidEndSepPunct{\mcitedefaultmidpunct}
{\mcitedefaultendpunct}{\mcitedefaultseppunct}\relax
\EndOfBibitem
\bibitem{CMS:2017hfy}
CMS collaboration, A.~M. Sirunyan {\em et~al.}, \ifthenelse{\boolean{articletitles}}{\emph{{Search for the $X(5568)$ state decaying into $B_s^0 \pi^\pm$ in proton-proton collisions at $\sqs = 8$\tev}}, }{}\href{https://doi.org/10.1103/PhysRevLett.120.202005}{Phys.\ Rev.\ Lett.\  \textbf{120} (2018) 202005}, \href{http://arxiv.org/abs/1712.06144}{{\normalfont\ttfamily arXiv:1712.06144}}\relax
\mciteBstWouldAddEndPuncttrue
\mciteSetBstMidEndSepPunct{\mcitedefaultmidpunct}
{\mcitedefaultendpunct}{\mcitedefaultseppunct}\relax
\EndOfBibitem
\bibitem{CDF:2017dwr}
CDF collaboration, T.~Aaltonen {\em et~al.}, \ifthenelse{\boolean{articletitles}}{\emph{{Search for the exotic meson $X(5568)$ with the collider detector at Fermilab}}, }{}\href{https://doi.org/10.1103/PhysRevLett.120.202006}{Phys.\ Rev.\ Lett.\  \textbf{120} (2018) 202006}, \href{http://arxiv.org/abs/1712.09620}{{\normalfont\ttfamily arXiv:1712.09620}}\relax
\mciteBstWouldAddEndPuncttrue
\mciteSetBstMidEndSepPunct{\mcitedefaultmidpunct}
{\mcitedefaultendpunct}{\mcitedefaultseppunct}\relax
\EndOfBibitem
\bibitem{ATLAS:2018udc}
ATLAS collaboration, M.~Aaboud {\em et~al.}, \ifthenelse{\boolean{articletitles}}{\emph{{Search for a structure in the $B^0_s \pi^\pm$ invariant mass spectrum with the ATLAS experiment}}, }{}\href{https://doi.org/10.1103/PhysRevLett.120.202007}{Phys.\ Rev.\ Lett.\  \textbf{120} (2018) 202007}, \href{http://arxiv.org/abs/1802.01840}{{\normalfont\ttfamily arXiv:1802.01840}}\relax
\mciteBstWouldAddEndPuncttrue
\mciteSetBstMidEndSepPunct{\mcitedefaultmidpunct}
{\mcitedefaultendpunct}{\mcitedefaultseppunct}\relax
\EndOfBibitem
\bibitem{LHCb-DP-2008-001}
LHCb collaboration, A.~A. Alves~Jr.\ {\em et~al.}, \ifthenelse{\boolean{articletitles}}{\emph{{The \lhcb detector at the LHC}}, }{}\href{https://doi.org/10.1088/1748-0221/3/08/S08005}{JINST \textbf{3} (2008) S08005}\relax
\mciteBstWouldAddEndPuncttrue
\mciteSetBstMidEndSepPunct{\mcitedefaultmidpunct}
{\mcitedefaultendpunct}{\mcitedefaultseppunct}\relax
\EndOfBibitem
\bibitem{LHCb-DP-2014-002}
LHCb collaboration, R.~Aaij {\em et~al.}, \ifthenelse{\boolean{articletitles}}{\emph{{LHCb detector performance}}, }{}\href{https://doi.org/10.1142/S0217751X15300227}{Int.\ J.\ Mod.\ Phys.\  \textbf{A30} (2015) 1530022}, \href{http://arxiv.org/abs/1412.6352}{{\normalfont\ttfamily arXiv:1412.6352}}\relax
\mciteBstWouldAddEndPuncttrue
\mciteSetBstMidEndSepPunct{\mcitedefaultmidpunct}
{\mcitedefaultendpunct}{\mcitedefaultseppunct}\relax
\EndOfBibitem
\bibitem{AdaBoost}
Y.~Freund and R.~E. Schapire, \ifthenelse{\boolean{articletitles}}{\emph{A decision-theoretic generalization of on-line learning and an application to boosting}, }{}\href{https://doi.org/10.1006/jcss.1997.1504}{J.\ Comput.\ Syst.\ Sci.\  \textbf{55} (1997) 119}\relax
\mciteBstWouldAddEndPuncttrue
\mciteSetBstMidEndSepPunct{\mcitedefaultmidpunct}
{\mcitedefaultendpunct}{\mcitedefaultseppunct}\relax
\EndOfBibitem
\bibitem{Punzi:2003bu}
G.~Punzi, \ifthenelse{\boolean{articletitles}}{\emph{{Sensitivity of searches for new signals and its optimization}}, }{}eConf \textbf{C030908} (2003) MODT002, \href{http://arxiv.org/abs/physics/0308063}{{\normalfont\ttfamily arXiv:physics/0308063}}\relax
\mciteBstWouldAddEndPuncttrue
\mciteSetBstMidEndSepPunct{\mcitedefaultmidpunct}
{\mcitedefaultendpunct}{\mcitedefaultseppunct}\relax
\EndOfBibitem
\bibitem{LHCb-TDR-001}
LHCb collaboration, \ifthenelse{\boolean{articletitles}}{\emph{{LHCb magnet: Technical Design Report}}, }{} \href{https://cds.cern.ch/search?p=CERN-LHCC-2000-007&f=reportnumber&action_search=Search&c=LHCb} {CERN-LHCC-2000-007}, 2000\relax
\mciteBstWouldAddEndPuncttrue
\mciteSetBstMidEndSepPunct{\mcitedefaultmidpunct}
{\mcitedefaultendpunct}{\mcitedefaultseppunct}\relax
\EndOfBibitem
\bibitem{Hulsbergen:2005pu}
W.~D. Hulsbergen, \ifthenelse{\boolean{articletitles}}{\emph{{Decay chain fitting with a Kalman filter}}, }{}\href{https://doi.org/10.1016/j.nima.2005.06.078}{Nucl.\ Instrum.\ Meth.\  \textbf{A552} (2005) 566}, \href{http://arxiv.org/abs/physics/0503191}{{\normalfont\ttfamily arXiv:physics/0503191}}\relax
\mciteBstWouldAddEndPuncttrue
\mciteSetBstMidEndSepPunct{\mcitedefaultmidpunct}
{\mcitedefaultendpunct}{\mcitedefaultseppunct}\relax
\EndOfBibitem
\bibitem{LHCb-PAPER-2011-019}
LHCb collaboration, R.~Aaij {\em et~al.}, \ifthenelse{\boolean{articletitles}}{\emph{{Measurement of the cross-section ratio $\sigma(\chictwo)/\sigma(\chicone)$ for prompt $\chi_c$ production at \mbox{$\sqs=$7~\tev}}}, }{}\href{https://doi.org/10.1016/j.physletb.2012.06.077}{Phys.\ Lett.\  \textbf{B714} (2012) 215}, \href{http://arxiv.org/abs/1202.1080}{{\normalfont\ttfamily arXiv:1202.1080}}\relax
\mciteBstWouldAddEndPuncttrue
\mciteSetBstMidEndSepPunct{\mcitedefaultmidpunct}
{\mcitedefaultendpunct}{\mcitedefaultseppunct}\relax
\EndOfBibitem
\bibitem{LHCb-PAPER-2013-028}
LHCb collaboration, R.~Aaij {\em et~al.}, \ifthenelse{\boolean{articletitles}}{\emph{{Measurement of the relative rate of prompt $\chiczero$, $\chicone$ and $\chictwo$ production at \mbox{$\sqs=$7~\tev}}}, }{}\href{https://doi.org/10.1007/JHEP10(2013)115}{JHEP \textbf{10} (2013) 115}, \href{http://arxiv.org/abs/1307.4285}{{\normalfont\ttfamily arXiv:1307.4285}}\relax
\mciteBstWouldAddEndPuncttrue
\mciteSetBstMidEndSepPunct{\mcitedefaultmidpunct}
{\mcitedefaultendpunct}{\mcitedefaultseppunct}\relax
\EndOfBibitem
\bibitem{LHCb-PAPER-2018-032}
LHCb collaboration, R.~Aaij {\em et~al.}, \ifthenelse{\boolean{articletitles}}{\emph{{Observation of two resonances in the $\Lb\pipm$ systems and precise measurement of $\Sigmabpm$ and $\Sigmares_b^{\ast\pm}$ properties}}, }{}\href{https://doi.org/10.1103/PhysRevLett.122.012001}{Phys.\ ~Rev.\ ~Lett.\  \textbf{122} (2019) 012001}, \href{http://arxiv.org/abs/1809.07752}{{\normalfont\ttfamily arXiv:1809.07752}}\relax
\mciteBstWouldAddEndPuncttrue
\mciteSetBstMidEndSepPunct{\mcitedefaultmidpunct}
{\mcitedefaultendpunct}{\mcitedefaultseppunct}\relax
\EndOfBibitem
\bibitem{Kolmogorov1933}
A.~N. Kolmogorov, \ifthenelse{\boolean{articletitles}}{\emph{{Sulla determinazione empirica di una legge di distribuzione}}, }{}Giornale dell'Istituto Italiano degli Attuari \textbf{4} (1933) 83\relax
\mciteBstWouldAddEndPuncttrue
\mciteSetBstMidEndSepPunct{\mcitedefaultmidpunct}
{\mcitedefaultendpunct}{\mcitedefaultseppunct}\relax
\EndOfBibitem
\bibitem{DAgostini:1999gfj}
G.~D'Agostini,  \ifthenelse{\boolean{articletitles}}{\emph{{Bayesian reasoning in high-energy physics: Principles and applications}}}{}, \href{https://doi.org/10.5170/CERN-1999-003}{CERN-1999-003}, CERN, 1999\relax
\mciteBstWouldAddEndPuncttrue
\mciteSetBstMidEndSepPunct{\mcitedefaultmidpunct}
{\mcitedefaultendpunct}{\mcitedefaultseppunct}\relax
\EndOfBibitem
\bibitem{Wilks:1938dza}
S.~S. Wilks, \ifthenelse{\boolean{articletitles}}{\emph{{The large-sample distribution of the likelihood ratio for testing composite hypotheses}}, }{}\href{https://doi.org/10.1214/aoms/1177732360}{Ann.\ Math.\ Stat.\  \textbf{9} (1938) 60}\relax
\mciteBstWouldAddEndPuncttrue
\mciteSetBstMidEndSepPunct{\mcitedefaultmidpunct}
{\mcitedefaultendpunct}{\mcitedefaultseppunct}\relax
\EndOfBibitem
\bibitem{Ebert:2009ua}
D.~Ebert, R.~N. Faustov, and V.~O. Galkin, \ifthenelse{\boolean{articletitles}}{\emph{{Heavy-light meson spectroscopy and Regge trajectories in the relativistic quark model}}, }{}\href{https://doi.org/10.1140/epjc/s10052-010-1233-6}{Eur.\ Phys.\ J.\  \textbf{C66} (2010) 197}, \href{http://arxiv.org/abs/0910.5612}{{\normalfont\ttfamily arXiv:0910.5612}}\relax
\mciteBstWouldAddEndPuncttrue
\mciteSetBstMidEndSepPunct{\mcitedefaultmidpunct}
{\mcitedefaultendpunct}{\mcitedefaultseppunct}\relax
\EndOfBibitem
\bibitem{Sun:2014wea}
Y.~Sun {\em et~al.}, \ifthenelse{\boolean{articletitles}}{\emph{{Higher bottom and bottom-strange mesons}}, }{}\href{https://doi.org/10.1103/PhysRevD.89.054026}{Phys.\ Rev.\  \textbf{D89} (2014) 054026}, \href{http://arxiv.org/abs/1401.1595}{{\normalfont\ttfamily arXiv:1401.1595}}\relax
\mciteBstWouldAddEndPuncttrue
\mciteSetBstMidEndSepPunct{\mcitedefaultmidpunct}
{\mcitedefaultendpunct}{\mcitedefaultseppunct}\relax
\EndOfBibitem
\bibitem{Godfrey:2016nwn}
S.~Godfrey, K.~Moats, and E.~S. Swanson, \ifthenelse{\boolean{articletitles}}{\emph{{$B$ and $B_s$ meson spectroscopy}}, }{}\href{https://doi.org/10.1103/PhysRevD.94.054025}{Phys.\ Rev.\  \textbf{D94} (2016) 054025}, \href{http://arxiv.org/abs/1607.02169}{{\normalfont\ttfamily arXiv:1607.02169}}\relax
\mciteBstWouldAddEndPuncttrue
\mciteSetBstMidEndSepPunct{\mcitedefaultmidpunct}
{\mcitedefaultendpunct}{\mcitedefaultseppunct}\relax
\EndOfBibitem
\bibitem{Lu:2016bbk}
Q.-F. L{\"u} {\em et~al.}, \ifthenelse{\boolean{articletitles}}{\emph{{Excited bottom and bottom-strange mesons in the quark model}}, }{}\href{https://doi.org/10.1103/PhysRevD.94.074012}{Phys.\ Rev.\  \textbf{D94} (2016) 074012}, \href{http://arxiv.org/abs/1607.02812}{{\normalfont\ttfamily arXiv:1607.02812}}\relax
\mciteBstWouldAddEndPuncttrue
\mciteSetBstMidEndSepPunct{\mcitedefaultmidpunct}
{\mcitedefaultendpunct}{\mcitedefaultseppunct}\relax
\EndOfBibitem
\bibitem{CMS:2025byz}
CMS collaboration, A.~Hayrapetyan {\em et~al.}, \ifthenelse{\boolean{articletitles}}{\emph{{First exclusive reconstruction of the $B^{*+}$, $B^{*0}$, and $B_s^{*0}$ mesons and precise measurement of their masses}}, }{}\href{https://doi.org/10.1103/njf9-4zfv}{Phys.\ Rev.\ Lett.\  \textbf{136} (2026) 031902}, \href{http://arxiv.org/abs/2508.05820}{{\normalfont\ttfamily arXiv:2508.05820}}\relax
\mciteBstWouldAddEndPuncttrue
\mciteSetBstMidEndSepPunct{\mcitedefaultmidpunct}
{\mcitedefaultendpunct}{\mcitedefaultseppunct}\relax
\EndOfBibitem
\bibitem{BaBar:2006eep}
BaBar collaboration, B.~Aubert {\em et~al.}, \ifthenelse{\boolean{articletitles}}{\emph{{A study of the $D^*_{sJ}(2317)^+$ and $D_{sJ}(2460)^+$ mesons in inclusive $\cquark \cquarkbar$ production near $\sqs = 10.6\gev$}}, }{}\href{https://doi.org/10.1103/PhysRevD.74.032007}{Phys.\ Rev.\  \textbf{D74} (2006) 032007}, \href{http://arxiv.org/abs/hep-ex/0604030}{{\normalfont\ttfamily arXiv:hep-ex/0604030}}\relax
\mciteBstWouldAddEndPuncttrue
\mciteSetBstMidEndSepPunct{\mcitedefaultmidpunct}
{\mcitedefaultendpunct}{\mcitedefaultseppunct}\relax
\EndOfBibitem
\bibitem{LHCb-PAPER-2026-019}
LHCb collaboration, R.~Aaij {\em et~al.}, \ifthenelse{\boolean{articletitles}}{\emph{{Study of muon-tagged $D_{s1}^+ \to D_s^+\pi^+\pi^-$ decays}}, }{}\href{http://arxiv.org/abs/2608.10826}{{\normalfont\ttfamily arXiv:2608.10826}}, {Submitted to JHEP}\relax
\mciteBstWouldAddEndPuncttrue
\mciteSetBstMidEndSepPunct{\mcitedefaultmidpunct}
{\mcitedefaultendpunct}{\mcitedefaultseppunct}\relax
\EndOfBibitem
\bibitem{BESIII:2025ksu}
BESIII collaboration, M.~Ablikim {\em et~al.}, \ifthenelse{\boolean{articletitles}}{\emph{{Precision measurement of $\Dss-\Ds$ mass difference}}, }{}\href{https://doi.org/https://doi.org/10.1016/j.scib.2026.04.059}{Science Bulletin \textbf{71} (2026) 2675}, \href{http://arxiv.org/abs/2510.20330}{{\normalfont\ttfamily arXiv:2510.20330}}\relax
\mciteBstWouldAddEndPuncttrue
\mciteSetBstMidEndSepPunct{\mcitedefaultmidpunct}
{\mcitedefaultendpunct}{\mcitedefaultseppunct}\relax
\EndOfBibitem
\bibitem{Nowak:1992um}
M.~A. Nowak, M.~Rho, and I.~Zahed, \ifthenelse{\boolean{articletitles}}{\emph{{Chiral effective action with heavy quark symmetry}}, }{}\href{https://doi.org/10.1103/PhysRevD.48.4370}{Phys.\ Rev.\  \textbf{D48} (1993) 4370}, \href{http://arxiv.org/abs/hep-ph/9209272}{{\normalfont\ttfamily arXiv:hep-ph/9209272}}\relax
\mciteBstWouldAddEndPuncttrue
\mciteSetBstMidEndSepPunct{\mcitedefaultmidpunct}
{\mcitedefaultendpunct}{\mcitedefaultseppunct}\relax
\EndOfBibitem
\bibitem{Bardeen:1993ae}
W.~A. Bardeen and C.~T. Hill, \ifthenelse{\boolean{articletitles}}{\emph{{Chiral dynamics and heavy quark symmetry in a solvable toy field theoretic model}}, }{}\href{https://doi.org/10.1103/PhysRevD.49.409}{Phys.\ Rev.\  \textbf{D49} (1994) 409}, \href{http://arxiv.org/abs/hep-ph/9304265}{{\normalfont\ttfamily arXiv:hep-ph/9304265}}\relax
\mciteBstWouldAddEndPuncttrue
\mciteSetBstMidEndSepPunct{\mcitedefaultmidpunct}
{\mcitedefaultendpunct}{\mcitedefaultseppunct}\relax
\EndOfBibitem
\bibitem{Melic:2025uha}
B.~Meli{\'c} and M.~Reboud, \ifthenelse{\boolean{articletitles}}{\emph{{Form factors and phenomenology of B$_{(s)}$ and D$_{(s)}$ semileptonic decays to {\ensuremath{\eta}} and {\ensuremath{\eta}}$^{\prime}$}}, }{}\href{https://doi.org/10.1007/JHEP10(2025)025}{JHEP \textbf{10} (2025) 025}, \href{http://arxiv.org/abs/2507.13734}{{\normalfont\ttfamily arXiv:2507.13734}}\relax
\mciteBstWouldAddEndPuncttrue
\mciteSetBstMidEndSepPunct{\mcitedefaultmidpunct}
{\mcitedefaultendpunct}{\mcitedefaultseppunct}\relax
\EndOfBibitem
\bibitem{Bharucha:2010im}
A.~Bharucha, T.~Feldmann, and M.~Wick, \ifthenelse{\boolean{articletitles}}{\emph{{Theoretical and phenomenological constraints on form factors for radiative and semi-leptonic $B$-meson decays}}, }{}\href{https://doi.org/10.1007/JHEP09(2010)090}{JHEP \textbf{09} (2010) 090}, \href{http://arxiv.org/abs/1004.3249}{{\normalfont\ttfamily arXiv:1004.3249}}\relax
\mciteBstWouldAddEndPuncttrue
\mciteSetBstMidEndSepPunct{\mcitedefaultmidpunct}
{\mcitedefaultendpunct}{\mcitedefaultseppunct}\relax
\EndOfBibitem
\bibitem{Biswas:2025drz}
A.~Biswas, N.~Gubernari, J.~Matias, and G.~Tetlalmatzi-Xolocotzi, \ifthenelse{\boolean{articletitles}}{\emph{{Impact on $L$-observables of a new combined analysis of $B_{d,s} \to K^{(*)}$ form factors}}, }{}\href{https://doi.org/10.1007/JHEP09(2025)188}{JHEP \textbf{09} (2025) 188}, \href{http://arxiv.org/abs/2506.12478}{{\normalfont\ttfamily arXiv:2506.12478}}\relax
\mciteBstWouldAddEndPuncttrue
\mciteSetBstMidEndSepPunct{\mcitedefaultmidpunct}
{\mcitedefaultendpunct}{\mcitedefaultseppunct}\relax
\EndOfBibitem
\bibitem{Savage:1990di}
M.~J. Savage and M.~B. Wise, \ifthenelse{\boolean{articletitles}}{\emph{{Spectrum of baryons with two heavy quarks}}, }{}\href{https://doi.org/10.1016/0370-2693(90)90035-5}{Phys.\ Lett.\  \textbf{B248} (1990) 177}\relax
\mciteBstWouldAddEndPuncttrue
\mciteSetBstMidEndSepPunct{\mcitedefaultmidpunct}
{\mcitedefaultendpunct}{\mcitedefaultseppunct}\relax
\EndOfBibitem
\bibitem{LHCb-PAPER-2017-018}
LHCb collaboration, R.~Aaij {\em et~al.}, \ifthenelse{\boolean{articletitles}}{\emph{{Observation of the doubly charmed baryon \Xiccpp}}, }{}\href{https://doi.org/10.1103/PhysRevLett.119.112001}{Phys.\ ~Rev.\ ~Lett.\  \textbf{119} (2017) 112001}, \href{http://arxiv.org/abs/1707.01621}{{\normalfont\ttfamily arXiv:1707.01621}}\relax
\mciteBstWouldAddEndPuncttrue
\mciteSetBstMidEndSepPunct{\mcitedefaultmidpunct}
{\mcitedefaultendpunct}{\mcitedefaultseppunct}\relax
\EndOfBibitem
\bibitem{LHCb-PAPER-2026-009}
LHCb collaboration, R.~Aaij {\em et~al.}, \ifthenelse{\boolean{articletitles}}{\emph{{Observation of the doubly charmed baryon \Xiccp with the LHCb Run 3 detector}}, }{}\href{https://doi.org/10.1103/dmv6-7gdv}{Phys.\ ~Rev.\ ~Lett.\  \textbf{137} (2026) 021902}, \href{http://arxiv.org/abs/2603.28456}{{\normalfont\ttfamily arXiv:2603.28456}}\relax
\mciteBstWouldAddEndPuncttrue
\mciteSetBstMidEndSepPunct{\mcitedefaultmidpunct}
{\mcitedefaultendpunct}{\mcitedefaultseppunct}\relax
\EndOfBibitem
\bibitem{LHCb-PAPER-2026-022}
LHCb collaboration, R.~Aaij {\em et~al.}, \ifthenelse{\boolean{articletitles}}{\emph{{Observation of the doubly charmed baryon $\Omegares_{cc}^+$}}, }{}\href{http://arxiv.org/abs/2609.21921}{{\normalfont\ttfamily arXiv:2609.21921}}, {submitted to Phys.~Rev.~Lett.}\relax
\mciteBstWouldAddEndPunctfalse
\mciteSetBstMidEndSepPunct{\mcitedefaultmidpunct}
{}{\mcitedefaultseppunct}\relax
\EndOfBibitem
\bibitem{LHCb-TDR-012}
LHCb collaboration, \ifthenelse{\boolean{articletitles}}{\emph{{Framework TDR for the LHCb Upgrade: Technical Design Report}}, }{} \href{https://cds.cern.ch/search?p=CERN-LHCC-2012-007&f=reportnumber&action_search=Search&c=LHCb} {CERN-LHCC-2012-007}, 2012\relax
\mciteBstWouldAddEndPuncttrue
\mciteSetBstMidEndSepPunct{\mcitedefaultmidpunct}
{\mcitedefaultendpunct}{\mcitedefaultseppunct}\relax
\EndOfBibitem
\bibitem{LHCb-TDR-002}
LHCb collaboration, \ifthenelse{\boolean{articletitles}}{\emph{{LHCb calorimeters: Technical Design Report}}, }{} \href{https://cds.cern.ch/search?p=CERN-LHCC-2000-036&f=reportnumber&action_search=Search&c=LHCb} {CERN-LHCC-2000-036}, 2000\relax
\mciteBstWouldAddEndPuncttrue
\mciteSetBstMidEndSepPunct{\mcitedefaultmidpunct}
{\mcitedefaultendpunct}{\mcitedefaultseppunct}\relax
\EndOfBibitem
\bibitem{Belyaev:2014zga}
I.~M. Belyaev, D.~Y. Golubkov, V.~Y. Egorychev, and D.~V. Savrina, \ifthenelse{\boolean{articletitles}}{\emph{{Calibration of the LHCb electromagnetic calorimeter using the technique of neutral pion invariant mass reconstruction}}, }{}\href{https://doi.org/10.1134/S0020441213060171}{Instrum.\ Exp.\ Tech.\  \textbf{57} (2014) 33}\relax
\mciteBstWouldAddEndPuncttrue
\mciteSetBstMidEndSepPunct{\mcitedefaultmidpunct}
{\mcitedefaultendpunct}{\mcitedefaultseppunct}\relax
\EndOfBibitem
\bibitem{LHCb-DP-2012-004}
R.~Aaij {\em et~al.}, \ifthenelse{\boolean{articletitles}}{\emph{{The \lhcb trigger and its performance in 2011}}, }{}\href{https://doi.org/10.1088/1748-0221/8/04/P04022}{JINST \textbf{8} (2013) P04022}, \href{http://arxiv.org/abs/1211.3055}{{\normalfont\ttfamily arXiv:1211.3055}}\relax
\mciteBstWouldAddEndPuncttrue
\mciteSetBstMidEndSepPunct{\mcitedefaultmidpunct}
{\mcitedefaultendpunct}{\mcitedefaultseppunct}\relax
\EndOfBibitem
\bibitem{LHCb-DP-2019-001}
R.~Aaij {\em et~al.}, \ifthenelse{\boolean{articletitles}}{\emph{{Design and performance of the LHCb trigger and full real-time reconstruction in Run 2 of the LHC}}, }{}\href{https://doi.org/10.1088/1748-0221/14/04/P04013}{JINST \textbf{14} (2019) P04013}, \href{http://arxiv.org/abs/1812.10790}{{\normalfont\ttfamily arXiv:1812.10790}}\relax
\mciteBstWouldAddEndPuncttrue
\mciteSetBstMidEndSepPunct{\mcitedefaultmidpunct}
{\mcitedefaultendpunct}{\mcitedefaultseppunct}\relax
\EndOfBibitem
\bibitem{BBDT}
V.~V. Gligorov and M.~Williams, \ifthenelse{\boolean{articletitles}}{\emph{{Efficient, reliable and fast high-level triggering using a bonsai boosted decision tree}}, }{}\href{https://doi.org/10.1088/1748-0221/8/02/P02013}{JINST \textbf{8} (2013) P02013}, \href{http://arxiv.org/abs/1210.6861}{{\normalfont\ttfamily arXiv:1210.6861}}\relax
\mciteBstWouldAddEndPuncttrue
\mciteSetBstMidEndSepPunct{\mcitedefaultmidpunct}
{\mcitedefaultendpunct}{\mcitedefaultseppunct}\relax
\EndOfBibitem
\bibitem{LHCb-PROC-2015-018}
T.~Likhomanenko {\em et~al.}, \ifthenelse{\boolean{articletitles}}{\emph{{LHCb topological trigger reoptimization}}, }{}\href{https://doi.org/10.1088/1742-6596/664/8/082025}{J.\ Phys.\ Conf.\ Ser.\  \textbf{664} (2015) 082025}, \href{http://arxiv.org/abs/1510.00572}{{\normalfont\ttfamily arXiv:1510.00572}}\relax
\mciteBstWouldAddEndPuncttrue
\mciteSetBstMidEndSepPunct{\mcitedefaultmidpunct}
{\mcitedefaultendpunct}{\mcitedefaultseppunct}\relax
\EndOfBibitem
\bibitem{Sjostrand:2007gs}
T.~Sj\"{o}strand, S.~Mrenna, and P.~Skands, \ifthenelse{\boolean{articletitles}}{\emph{{A brief introduction to PYTHIA 8.1}}, }{}\href{https://doi.org/10.1016/j.cpc.2008.01.036}{Comput.\ Phys.\ Commun.\  \textbf{178} (2008) 852}, \href{http://arxiv.org/abs/0710.3820}{{\normalfont\ttfamily arXiv:0710.3820}}\relax
\mciteBstWouldAddEndPuncttrue
\mciteSetBstMidEndSepPunct{\mcitedefaultmidpunct}
{\mcitedefaultendpunct}{\mcitedefaultseppunct}\relax
\EndOfBibitem
\bibitem{LHCb-PROC-2010-056}
I.~Belyaev {\em et~al.}, \ifthenelse{\boolean{articletitles}}{\emph{{Handling of the generation of primary events in Gauss, the LHCb simulation framework}}, }{}\href{https://doi.org/10.1088/1742-6596/331/3/032047}{J.\ Phys.\ Conf.\ Ser.\  \textbf{331} (2011) 032047}\relax
\mciteBstWouldAddEndPuncttrue
\mciteSetBstMidEndSepPunct{\mcitedefaultmidpunct}
{\mcitedefaultendpunct}{\mcitedefaultseppunct}\relax
\EndOfBibitem
\bibitem{Lange:2001uf}
D.~J. Lange, \ifthenelse{\boolean{articletitles}}{\emph{{The EvtGen particle decay simulation package}}, }{}\href{https://doi.org/10.1016/S0168-9002(01)00089-4}{Nucl.\ Instrum.\ Meth.\  \textbf{A462} (2001) 152}\relax
\mciteBstWouldAddEndPuncttrue
\mciteSetBstMidEndSepPunct{\mcitedefaultmidpunct}
{\mcitedefaultendpunct}{\mcitedefaultseppunct}\relax
\EndOfBibitem
\bibitem{davidson2015photos}
N.~Davidson, T.~Przedzinski, and Z.~Was, \ifthenelse{\boolean{articletitles}}{\emph{{PHOTOS interface in C++: Technical and physics documentation}}, }{}\href{https://doi.org/https://doi.org/10.1016/j.cpc.2015.09.013}{Comput.\ Phys.\ Commun.\  \textbf{199} (2016) 86}, \href{http://arxiv.org/abs/1011.0937}{{\normalfont\ttfamily arXiv:1011.0937}}\relax
\mciteBstWouldAddEndPuncttrue
\mciteSetBstMidEndSepPunct{\mcitedefaultmidpunct}
{\mcitedefaultendpunct}{\mcitedefaultseppunct}\relax
\EndOfBibitem
\bibitem{Allison:2006ve}
Geant4 collaboration, J.~Allison {\em et~al.}, \ifthenelse{\boolean{articletitles}}{\emph{{Geant4 developments and applications}}, }{}\href{https://doi.org/10.1109/TNS.2006.869826}{IEEE Trans.\ Nucl.\ Sci.\  \textbf{53} (2006) 270}\relax
\mciteBstWouldAddEndPuncttrue
\mciteSetBstMidEndSepPunct{\mcitedefaultmidpunct}
{\mcitedefaultendpunct}{\mcitedefaultseppunct}\relax
\EndOfBibitem
\bibitem{Agostinelli:2002hh}
Geant4 collaboration, S.~Agostinelli {\em et~al.}, \ifthenelse{\boolean{articletitles}}{\emph{{Geant4: A simulation toolkit}}, }{}\href{https://doi.org/10.1016/S0168-9002(03)01368-8}{Nucl.\ Instrum.\ Meth.\  \textbf{A506} (2003) 250}\relax
\mciteBstWouldAddEndPuncttrue
\mciteSetBstMidEndSepPunct{\mcitedefaultmidpunct}
{\mcitedefaultendpunct}{\mcitedefaultseppunct}\relax
\EndOfBibitem
\bibitem{LHCb-PROC-2011-006}
M.~Clemencic {\em et~al.}, \ifthenelse{\boolean{articletitles}}{\emph{{The \lhcb simulation application, Gauss: Design, evolution and experience}}, }{}\href{https://doi.org/10.1088/1742-6596/331/3/032023}{J.\ Phys.\ Conf.\ Ser.\  \textbf{331} (2011) 032023}\relax
\mciteBstWouldAddEndPuncttrue
\mciteSetBstMidEndSepPunct{\mcitedefaultmidpunct}
{\mcitedefaultendpunct}{\mcitedefaultseppunct}\relax
\EndOfBibitem
\bibitem{LHCb-DP-2013-002}
LHCb collaboration, R.~Aaij {\em et~al.}, \ifthenelse{\boolean{articletitles}}{\emph{{Measurement of the track reconstruction efficiency at LHCb}}, }{}\href{https://doi.org/10.1088/1748-0221/10/02/P02007}{JINST \textbf{10} (2015) P02007}, \href{http://arxiv.org/abs/1408.1251}{{\normalfont\ttfamily arXiv:1408.1251}}\relax
\mciteBstWouldAddEndPuncttrue
\mciteSetBstMidEndSepPunct{\mcitedefaultmidpunct}
{\mcitedefaultendpunct}{\mcitedefaultseppunct}\relax
\EndOfBibitem
\bibitem{LHCb-DP-2013-001}
F.~Archilli {\em et~al.}, \ifthenelse{\boolean{articletitles}}{\emph{{Performance of the muon identification at LHCb}}, }{}\href{https://doi.org/10.1088/1748-0221/8/10/P10020}{JINST \textbf{8} (2013) P10020}, \href{http://arxiv.org/abs/1306.0249}{{\normalfont\ttfamily arXiv:1306.0249}}\relax
\mciteBstWouldAddEndPuncttrue
\mciteSetBstMidEndSepPunct{\mcitedefaultmidpunct}
{\mcitedefaultendpunct}{\mcitedefaultseppunct}\relax
\EndOfBibitem
\bibitem{Cowan:2016tnm}
G.~A. Cowan, D.~C. Craik, and M.~D. Needham, \ifthenelse{\boolean{articletitles}}{\emph{{RapidSim: an application for the fast simulation of heavy-quark hadron decays}}, }{}\href{https://doi.org/10.1016/j.cpc.2017.01.029}{Comput.\ Phys.\ Commun.\  \textbf{214} (2017) 239}, \href{http://arxiv.org/abs/1612.07489}{{\normalfont\ttfamily arXiv:1612.07489}}\relax
\mciteBstWouldAddEndPuncttrue
\mciteSetBstMidEndSepPunct{\mcitedefaultmidpunct}
{\mcitedefaultendpunct}{\mcitedefaultseppunct}\relax
\EndOfBibitem
\bibitem{XGBoosting}
T.~Chen and C.~Guestrin, \ifthenelse{\boolean{articletitles}}{\emph{{XGBoost}: A scalable tree boosting system}, }{} in {\em Proceedings of the 22nd ACM SIGKDD International Conference on Knowledge Discovery and Data Mining}, \href{https://doi.org/10.1145/2939672.2939785}{ KDD '16, (New York, NY, USA), 785--794, ACM, 2016}, \href{http://arxiv.org/abs/1603.02754}{{\normalfont\ttfamily arXiv:1603.02754}}\relax
\mciteBstWouldAddEndPuncttrue
\mciteSetBstMidEndSepPunct{\mcitedefaultmidpunct}
{\mcitedefaultendpunct}{\mcitedefaultseppunct}\relax
\EndOfBibitem
\bibitem{Skwarnicki:1986xj}
T.~Skwarnicki, {\em {A study of the radiative cascade transitions between the Upsilon-prime and Upsilon resonances}}, PhD thesis, Institute of Nuclear Physics, Krakow, 1986, {\href{http://inspirehep.net/record/230779/}{DESY-F31-86-02}}\relax
\mciteBstWouldAddEndPuncttrue
\mciteSetBstMidEndSepPunct{\mcitedefaultmidpunct}
{\mcitedefaultendpunct}{\mcitedefaultseppunct}\relax
\EndOfBibitem
\bibitem{Blatt:1952ije}
J.~M. Blatt and V.~F. Weisskopf, {\em {Theoretical nuclear physics}}, \href{https://doi.org/10.1007/978-1-4612-9959-2}{ Springer, New York, 1952}\relax
\mciteBstWouldAddEndPuncttrue
\mciteSetBstMidEndSepPunct{\mcitedefaultmidpunct}
{\mcitedefaultendpunct}{\mcitedefaultseppunct}\relax
\EndOfBibitem
\bibitem{LHCb-DP-2023-003}
LHCb collaboration, R.~Aaij {\em et~al.}, \ifthenelse{\boolean{articletitles}}{\emph{{Momentum scale calibration of the LHCb spectrometer}}, }{}\href{https://doi.org/10.1088/1748-0221/19/02/P02008}{JINST \textbf{19} (2024) P02008}, \href{http://arxiv.org/abs/2312.01772}{{\normalfont\ttfamily arXiv:2312.01772}}\relax
\mciteBstWouldAddEndPuncttrue
\mciteSetBstMidEndSepPunct{\mcitedefaultmidpunct}
{\mcitedefaultendpunct}{\mcitedefaultseppunct}\relax
\EndOfBibitem
\bibitem{LHCb-PAPER-2012-030}
LHCb collaboration, R.~Aaij {\em et~al.}, \ifthenelse{\boolean{articletitles}}{\emph{{First observation of the decay \mbox{\decay{B_{s2}^*(5840)^0}{B^{*+}\Km}} and studies of excited \Bs mesons}}, }{}\href{https://doi.org/10.1103/PhysRevLett.110.151803}{Phys.\ ~Rev.\ ~Lett.\  \textbf{110} (2013) 151803}, \href{http://arxiv.org/abs/1211.5994}{{\normalfont\ttfamily arXiv:1211.5994}}\relax
\mciteBstWouldAddEndPuncttrue
\mciteSetBstMidEndSepPunct{\mcitedefaultmidpunct}
{\mcitedefaultendpunct}{\mcitedefaultseppunct}\relax
\EndOfBibitem
\bibitem{LHCb-PAPER-2020-026}
LHCb collaboration, R.~Aaij {\em et~al.}, \ifthenelse{\boolean{articletitles}}{\emph{{Observation of new excited $\Bs$ states}}, }{}\href{https://doi.org/10.1140/epjc/s10052-021-09305-3}{Eur.\ Phys.\ J.\  \textbf{C81} (2021) 601}, \href{http://arxiv.org/abs/2010.15931}{{\normalfont\ttfamily arXiv:2010.15931}}\relax
\mciteBstWouldAddEndPuncttrue
\mciteSetBstMidEndSepPunct{\mcitedefaultmidpunct}
{\mcitedefaultendpunct}{\mcitedefaultseppunct}\relax
\EndOfBibitem
\end{mcitethebibliography}

\newpage
% LHCb collaboration author list
% Data extracted on September 13th, 2026 at 9:28am for paper reference LHCb-PAPER-2026-021
\centerline
{\large\bf LHCb collaboration}
\begin
{flushleft}
\small
R.~Aaij$^{39}$\lhcborcid{0000-0003-0533-1952},
M.~Abdelfatah$^{71}$,
A.S.W.~Abdelmotteleb$^{59}$\lhcborcid{0000-0001-7905-0542},
C.~Abellan~Beteta$^{53}$\lhcborcid{0009-0009-0869-6798},
F.~Abudin\'en$^{61}$\lhcborcid{0000-0002-6737-3528},
T.~Ackernley$^{63}$\lhcborcid{0000-0002-5951-3498},
A.A.~Adefisoye$^{71}$\lhcborcid{0000-0003-2448-1550},
B.~Adeva$^{49}$\lhcborcid{0000-0001-9756-3712},
M.~Adinolfi$^{57}$\lhcborcid{0000-0002-1326-1264},
P.~Adlarson$^{87,44}$\lhcborcid{0000-0001-6280-3851}
% ...

{\footnotesize \it

% ...
$^{39}$Nikhef National Institute for Subatomic Physics, Amsterdam, Netherlands\\
% ...
$^{44}$National Center for Nuclear Research (NCBJ), Warsaw, Poland\\
% ...
$^{49}$Instituto Galego de F{\'\i}sica de Altas Enerx{\'\i}as (IGFAE), Universidade de Santiago de Compostela, Santiago de Compostela, Spain\\
% ...
$^{53}$Physik-Institut, Universit{\"a}t Z{\"u}rich, Z{\"u}rich, Switzerland\\
% ...
$^{57}$H.H. Wills Physics Laboratory, University of Bristol, Bristol, United Kingdom\\
% ...
$^{59}$Department of Physics, University of Warwick, Coventry, United Kingdom\\
% ...
$^{61}$School of Physics and Astronomy, University of Edinburgh, Edinburgh, United Kingdom\\
$^{62}$School of Physics and Astronomy, University of Glasgow, Glasgow, United Kingdom\\
$^{63}$Oliver Lodge Laboratory, University of Liverpool, Liverpool, United Kingdom\\
% ...
$^{71}$Syracuse University, Syracuse, NY, United States\\
% ...
$^{87}$Department of Physics and Astronomy, Uppsala University, Uppsala, Sweden, associated to $^{62}$\\
% ...
}
\end{flushleft}

\end{document}